\documentclass[12pt]{article}

\usepackage{amssymb}
\usepackage{color}
\usepackage{epsfig,amssymb,amsfonts,amsmath,graphicx,dsfont,cite,xfrac}
\usepackage{authblk}
\usepackage{subcaption}

\usepackage[colorlinks=true,linkcolor=blue,citecolor=blue]{hyperref}

\newcommand{\be}{\begin{equation}}
\newcommand{\ee}{\end{equation}}
\newcommand{\bea}{\begin{eqnarray}}
\newcommand{\eea}{\end{eqnarray}}

\title{Time-evolution of quantum systems via a complex nonlinear Riccati equation II. Dissipative systems}

\author[1]{Hans Cruz}
\author[2]{Dieter Schuch}
\author[1]{Octavio Casta\~nos}
\author[3]{Oscar Rosas-Ortiz}

\affil[1]{\footnotesize Instituto de Ciencias Nucleares, Universidad Nacional Aut\'onoma de M\'exico, A. P. 70-543, 04510 M\'exico D.F., Mexico}

\affil[2]{\footnotesize Institut f\"ur Theoretische Physik, JW Goethe-Universit\"at Frankfurt am Main, Max-von-Laue-Str. 1, D-60438 Frankfurt am Main, Germany}

\affil[3]{\footnotesize Physics Department, Cinvestav, AP 14-740, 07000
M\'exico DF, Mexico}

\date{}
\begin{document}

\maketitle

\begin{abstract}
In our former contribution~\cite{CSCR}, we have shown the sensitivity to the choice of initial conditions in the evolution of Gaussian wave packets via the nonlinear Riccati equation. The formalism developed in the previous work is extended to effective approaches for the description of dissipative quantum systems. By means of simple examples we show the effects of the environment on the quantum uncertainties, correlation function, quantum energy contribution and tunnelling currents. We prove that the environmental parameter $\gamma$ is strongly related with the sensitivity to the choice of initial conditions.
\end{abstract}



\section{Introduction}

\label {section-1}

In part I~\cite{CSCR}, it has been shown for non-dissipative\footnote{Usually, ``dissipative'' is set against ``conservative'', this last describing, e.g., systems where the energy and therefore the Hamiltonian function is a constant of motion. However, in part I  also the parametric oscillator with $\omega=\omega(t)$ was considered, leading to a TD Hamiltonian ${H}(t)$, i.e., non-conservative, but without dissipation. To distinguish this type of systems included in part I from the dissipative ones, in part II we refer to the first type as ``non-dissipative" systems.} systems that the information about the dynamics of a quantum system that can be obtained from the time-dependent (TD) Schr\"odinger equation (SE) can equally well be obtained from the complex, quadratically nonlinear (NL), Riccati equation in the cases where the TDSE possesses exact analytic solutions in the form of Gaussian wave packets (WPs), i.e., for Hamiltonians that are at most quadratic or bilinear in position and momentum. As Gaussian functions are completely determined by their maximum and their width, this led to the problem of solving the equations of motion for these two parameters.   

The maximum followed, according to Ehrenfest's theorem, the classical equation of motion for the mean value of the position variable. The equation of motion for the WP's width was related with the complex TD coefficient of the quadratic term in the exponent of the Gaussian. This coefficient fulfilled a NL Riccati equation. Taking the square root of the inverse of the imaginary part of the variable fulfilling this Riccati equation as a new variable, this was directly proportional to the WP's width and fulfilled a (real) NL so-called Ermakov equation. This equation, together with the Newtonian equation for the WP's maximum allowed to obtain a dynamical invariant even for an oscillator with $\omega=\omega(t)$, i.e., in cases where the Hamiltonian ${H}(t)$ is no longer an invariant. 

Due to the nonlinearity of the Riccati equation, the sensitivity to the choice of the initial conditions has been shown in~\cite{CSCR} for the non-dissipative case. This and the consequences for physical properties of the system like ground state energy and tunnelling currents will become even more obvious in the dissipative case discussed in this paper. However, the first problem to be solved is to find an appropriate quantum mechanical description of the dissipative system compatible with the TDSE used in the non-dissipative case. This is problematic as dissipative effects cannot be included into the established classical Lagrangian or Hamiltonian formalism using canonical transformations and serving as a basis for the so-called canonical quantization. Therefore, in Section~\ref{section-2} some possibilities are shown to solve this problem. In order to preserve the canonical formalism and linear (but possibly explicitly TD) Hamiltonian operators after quantization, one has to allow for non-canonical (classical) or non-unitary (quantum mechanical) transformations. This has the disadvantage that physical and canonical variables have to be strictly distinguished and their interrelation has to be considered carefully.

The other possibility is to keep the physical meaning of position and momentum variables and the corresponding operators like in standard quantum mechanics, but this can lead to NL and non-Hermitian Hamiltonian operators, implying nonlinear Schr\"odinger equations (NLSEs).

The pleasant feature of these two types of approaches is that they can uniquely be linked via non-canonical (classical) or non-unitary (quantum mechanical) transformations, as is also shown in Section~\ref{section-2}.

In the cases of interest, i.e., Hamiltonians that are at most quadratic or bilinear in the canonical variables, also for dissipative quantum mechanical systems, exact analytic solutions with the form of Gaussian WPs exist. For the different approaches the equations of motion for the maximum and width of these Gaussians can be transformed consistently into each other. Therefore, it is sufficient to solve these equations for the case where the canonical variables are identical with the physical position and momentum, leading to the solution of a NLSE.  

However, formal advantages of the linear canonical transformations can also be exploited to find analogies between the treatment of the non-dissipative systems and the dissipative ones, particularly with respect to the choice of the initial conditions, what will be shown in Section~\ref{section-3}.  

In Section~\ref{section-4}, it is shown how the complex Riccati equation obtained from the NLSE can be linearized to a complex Newtonian equation including a linear velocity dependent friction term. Also the Feynman kernel (propagator) for the dissipative case can be obtained in analogy to the one without dissipation. As in our first contribution~\cite{CSCR}, in Section~\ref{section-5} the Wigner function is presented and it is shown that this quasi-distribution of probability, as in standard quantum theory, determines the probability densities in position and momentum space via the marginals. 

The explicit time-dependence of the position and momentum uncertainties, their correlations and uncertainty relations as well as the tunnelling currents and the quantum contributions to the energy, depending on the initial conditions, are given in Section~\ref{section-6}. It is shown that different initial values, e.g., no initial change in time of the width ($\dot{\alpha}_0=0$), or, no initial correlation of position and momentum uncertainties ($\sigma_{xp_0}=\frac{\hbar}{2}\alpha_0\left[\dot{\alpha}_0-\frac{\gamma}{2}\alpha_0\right]=0$) lead to different results and therefore different values for the physical properties of the system. This is shown for the damped free motion and the damped harmonic oscillator (HO) (with $\omega_0 > \frac{\gamma}{2}$, $\omega_0 = \frac{\gamma}{2}$ and $\omega_0 < \frac{\gamma}{2}$).

Finally, in Section~\ref{section-7} the results are summarized, some conclusions are drawn and perspectives given. 


\section{Effective description of dissipative systems }

\label{section-2}

Classical Hamiltonian mechanics and quantum mechanics describe isolated systems with reversible dynamics. However, realistic physical systems are always in contact with some kind of environment. This coupling usually introduces the phenomena of irreversibility and dissipation. How can this be taken into account in the formalism of classical (Hamiltonian or Lagrangian) mechanics and, particularly, in a quantum mechanical context? The fundamental classical as well as quantum mechanical equations of motion are invariant under time-reversal and the forces are assumed to be derived from the gradient of a potential, guaranteeing for time-independent potentials conservation of energy. In classical mechanics, the time-evolution can be traced back to canonical transformations, in quantum mechanics to unitary transformations.

Alternative approaches to include irreversibility and dissipation are:

\begin{enumerate}

\item Phenomenological equations like the Langevin or Fokker--Planck equations~\cite{Chandrasekhar, Weiss}.

\item System-plus-reservoir approaches~\cite{Weiss, vankampen, Caldeira-Leggett, Kossakowski, Lindblad}.

\item Modifications of the classical and quantum mechanical equations of motion, leading to non-canonical/non-unitary transformations or NL modifications of the SE~\cite{Gisin, Albrecht, Hasse, Schuch-Chung, elaf2010}.

\end{enumerate}  

In the phenomenological description where a Brownian motion-type situation is assumed, i.e., a macroscopic body moving in a viscous liquid (actually a many-body problem), the effect of the bath is described by adding a linear velocity-depending friction force and a stochastic fluctuating force (that vanishes on average) to the Newtonian equation for the observable macroscopic system. In this effective description of the system the equation of motion for the average position variable, in one dimension, as assumed throughout this paper without loss of generality, takes the form
\begin{equation}
m\ddot{x}+m\gamma\dot{x}+\frac{\partial V(x)}{\partial x}=0
\label{Lang-eq}
\end{equation}
with $\gamma$ the friction coefficient (\emph{damping constant}) and overdots denoting time-derivatives. As a consequence of the friction force $-m\gamma\dot{x}$, the velocity and energy of the system decay exponentially.

For a HO ($V(x) =\frac12 m \omega^2 x^2$), the friction coefficient $\gamma$ is usually referred to as {\em Ohmic damping} because Eq. (1) also describes an RLC circuit in series with $\gamma =R/L$, $\omega^{-2} =LC$ and $q=x/m$ the `charge' of the capacitor. In classical physics damping can be introduced without knowledge of the microscopic details of the bath. Although this is often described by a friction force that is linear in velocity, as in the Langevin equation (1), friction forces that are quadratic in velocity are also relevant for different systems (see, e.g.,~\cite{A,B,C}  and references quoted therein). 

A description of the same situation, not in terms of trajectories, but in terms of (classical) statistical distribution functions $\rho_{\tiny \mbox{cl}}(x,p,t)$ in phase space can be given in terms of a Fokker--Planck equation taking  into account irreversible diffusion terms. In particular, in position space this reduces to the Smoluchowski equation
\begin{equation}
\frac{\partial\rho_{\tiny \mbox{cl}}}{\partial t} +\frac{\partial}{\partial x} \bigl( \rho_{\tiny \mbox{cl}}\, \dot{x} \bigr) - D\frac{\partial^2\rho_{\tiny \mbox{cl}}}{\partial x^2}=0 \, ,
\end{equation}
with the diffusion coefficient $D$ that is usually connected with the temperature $T$ of the bath via the Einstein relation $D=\frac{kT}{m\gamma}$, where $k$ is Boltzmann's constant.

In the system-plus-reservoir approaches, the system of interest is coupled to an environment (consisting usually of a large number of HOs), where the system and environment together are considered as a closed Hamiltonian system. After averaging over the environmental degrees of freedom and some others procedures  (for details see, e.g.,~\cite{Weiss}) an equation of motion for the system of interest including a friction term, e.g., like the one in the Langevin equation, is obtained.

Starting point for system-plus-reservoir approaches is usually, on the classical level, the Liouville equation for the distribution function $\rho_{\tiny \mbox{cl}}(x,p,t)$ in phase space and, quantum mechanically, the equivalent von Neumann equation for the density operator in Liouville space. In both cases they are then decomposed (e.g. by some kind of projection operator) into a relevant part, describing the system of interest, and an irrelevant part, essentially describing the environment. Solving the equation formally for the irrelevant part and inserting this into the equation for the relevant one leads (after a Markovian approximation) to the so-called generalized master equation~\cite{vankampen} for the relevant part, describing the system that dissipates energy to the environment. Frequently, an approach by Caldeira and Leggett~\cite{Caldeira-Leggett} that is based on this method and modifications thereof are applied. Interestingly for our purpose, this approach can be linked directly to the quantized version of a modification of the classical canonical Lagrangian and Hamiltonian formalism~\cite{Caldirola, Kanai} that allows one to include linear velocity-dependent friction forces like the one in Eq.~(\ref{Lang-eq}). More details are given in Subsection~\ref{section-2.1}.

A different approach, also based on the generalized master equation of the system-plus-reservoir approach, assumes that a quantum system can always be described by a pure state, no matter if it is in contact with an environment or not~\cite{Gisin}. In analogy with the generalized master equations mentioned above, Gisin derived a corresponding equation of motion for this pure state, represented by a wave function. This leads to a NLSE with non-hermitian (complex) nonlinearity that still allows for normalizable solutions. However, the specific form of Gisin's nonlinearity and its physical interpretation are not fixed by its derivation. 

It turns out that a modified SE, based on a Smoluchowski equation for the probability density $\rho(x,t)=\Psi^\ast(x,t)\Psi(x,t)$, leads to a formally similar NLSE, but here with a fixed form of the (complex) nonlinearity and well-defined physical interpretation. This approach will be discussed in detail in Subsection~\ref{section-2.2}, as well as its unique connection with the canonical approaches described in Subsection~\ref{section-2.1}.


\subsection{Canonical (Unitary) description of dissipative systems }
\label{section-2.1}

One type of effective models that is significant for our purpose includes the models that stay within the canonical formalism, thus providing the possibility of obtaining a corresponding SE via canonical quantization. Attempts to obtain the friction force of Eq. (\ref{Lang-eq}) by simply adding a term to the potential $V(x)$ are neither in the Lagrangian nor in the Hamiltonian formalism successful. However, an approach by Caldirola~\cite{Caldirola} and Kanai~\cite{Kanai}, multiplying the Lagrangian function by a TD exponential factor leads to the desired result. The classical version starts with the explicitly TD Lagrangian (from here on $V(x) = \frac{1}{2} \, m \, \omega^2 x^2$)
\begin{equation}
\hat{L}_{\tiny \mbox{CK}}(\hat{x},\hat{p}) = \left[ \frac{m}{2}\dot{\hat{x}}^2-\frac{1}{2}m\omega^2 \hat{x}^2 \right]e^{\gamma t}\, ,
\end{equation}
leading, via the Euler--Lagrange equation, to the averaged Langevin equation (\ref{Lang-eq}). Here and in the following, canonical variables and corresponding Lagrangians/Hamiltonians are characterized by a hat ( $\hat{}$ ). This also applies in the quantized version to the canonical operators and wave functions.

In this case, the physical position variable $x$ and the canonical one, $\hat{x}$, coincide; however, the canonical momentum $\hat{p}$ is related with the physical one, $p$, (with $\hat{x}=x$) via
\begin{equation}
\frac{\partial}{\partial \dot{x}}\hat{L}_{\tiny \mbox{CK}}=m\dot{x}e^{\gamma t}=p e^{\gamma t}=\hat{p}\, .
\end{equation}
Note, that the transition from the physical variables $(x,p)$ to the canonical variables $(\hat{x}=x,\hat{p}=p e^{\gamma t})$ is a non-canonical transformation\footnote{The Jacobian determinant is different from one.}.

It is straightforward to obtain the corresponding Hamiltonian $\hat{H}_{\tiny \mbox{CK}}(\hat{x},\hat{p})$ in the form
\begin{equation}
\hat{H}_{\tiny \mbox{CK}}(x,\hat{p}) = e^{-\gamma t} \, \frac{\hat{p}^2}{2m}+ e^{\gamma t} \, \frac{m}{2} \, \omega^2 x^2\, ,
\label{H-CK}
\end{equation} 
which also supplies the correct equation of motion including the friction force.

Although this explicitly  TD Hamiltonian is not a constant of motion, from its quantized version an exact Ermakov invariant can be derived~\cite{elaf2010} even in the case where $\omega$ is TD i.e., $\omega=\omega(t)$. For this purpose, canonical quantization is used, i.e., the canonical momentum $\hat{p}$ is replaced by operator $\hat{p}_{\tiny \mbox{op}}=\frac{\hbar}{i}\frac{\partial}{\partial x}$. Applying the resulting (linear, but explicitly TD) Hamiltonian operator $\hat{H}_{\tiny \mbox{CK,op}}$ to the {\it canonical} wave function $\hat{\Psi}_{\tiny \mbox{CK}}(x,t)$ leads to the modified SE
\begin{equation}
i\hbar\frac{\partial}{\partial t}\hat{\Psi}_{\tiny \mbox{CK}}=\left[-\frac{\hbar^2}{2m}e^{-\gamma t}\frac{\partial^2}{\partial x^2}+e^{\gamma t} \, \frac{m}{2} \, \omega^2 x^2\right]\hat{\Psi}_{\tiny \mbox{CK}}\, .
\label{SE-CK}
\end{equation}
As in the cases discussed for the non-dissipative TDSE in part I, this equation also possesses exact Gaussian WP solutions of the form
\begin{equation}
\hat{\Psi}_{\tiny \mbox{CK}}(x,t)=\hat{N}_{\tiny \mbox{CK}}(t)\exp\left\{ i\left[ \hat{y}_{\tiny \mbox{CK}}(t)\tilde{x}^2+\frac{1}{\hbar}\langle \hat{p} \rangle_{\tiny \mbox{CK}}(t)\tilde{x}+\hat{K}_{\tiny \mbox{CK}}(t) \right] \right\}
\end{equation}
with the complex TD parameter $\hat{y}_{\tiny \mbox{CK}}(t)$ and $\tilde{x}=x-\langle x \rangle_{\tiny \mbox{CK}}(t)=x-\eta(t)$. The purely TD functions $\hat{N}_{\tiny \mbox{CK}}(t)$ and $\hat{K}_{\tiny \mbox{CK}}(t)$ are not relevant for the following discussion. Here, $\langle . . . \rangle_{\tiny \mbox{CK}}$ indicates that the mean value is calculated using the {\it canonical} wave function $\hat{\Psi}_{\tiny \mbox{CK}}(x,t)$. 

The equation of motion for the WP maximum is just the one for the classical trajectory including the friction force, i.e., in the notation of part I,
\begin{equation}
\ddot{\eta}+\gamma\,\dot{\eta}+\omega^2 \eta=0\, .
\label{Dis-New-Eq}
\end{equation}
The modified complex Riccati equation for the complex coefficient $\hat{y}_{\tiny \mbox{CK}}(t)$ of the quadratic term in the exponent, or, $\hat{C}_{\tiny \mbox{CK}}(t)=\frac{2\hbar}{m}\hat{y}_{\tiny \mbox{CK}}(t)$,
\begin{equation}
\dot{\hat{C}}_{\tiny \mbox{CK}}+e^{-\gamma t}\,\hat{C}^2_{\tiny \mbox{CK}}+e^{\gamma t}\,\omega^2=0\, ,
\label{Riccati-CK}
\end{equation}
can also be transformed, as in the non-dissipative case, using the definition of its imaginary part as\begin{equation}
\hat{C}_{\tiny \mbox{CK,I}} =\frac{\hbar}{2m\langle \tilde{x}^2 \rangle_{\tiny \mbox{CK}}}=\frac{1}{\hat{\alpha}^2_{\tiny \mbox{CK}}}\, ,
\end{equation}
with $\langle \tilde{x}^2 \rangle_{\tiny \mbox{CK}}=\langle x^2 \rangle_{\tiny \mbox{CK}}-\langle x \rangle^2_{\tiny \mbox{CK}}$, into a real Ermakov-type equation for $\hat{\alpha}_{\tiny \mbox{CK}}(t)$,
\begin{equation}
\ddot{\hat{\alpha}}_{\tiny \mbox{CK}}+\gamma\,\dot{\hat{\alpha}}_{\tiny \mbox{CK}}+\omega^2 \,\hat{\alpha}_{\tiny \mbox{CK}}=\frac{e^{-2\gamma t}}{\hat{\alpha}^3_{\tiny \mbox{CK}}}\,.
\end{equation}
This equation, together with Eq. (\ref{Dis-New-Eq}), forms a system of equations of motion, coupled  via $\omega$, that possesses an exact Ermakov-type invariant~\cite{elaf2010}, now given in the form
\begin{equation}
\hat{I}_{\tiny \mbox{CK}}=\frac{m}{2\hbar}\left[ \left( \dot{\eta} \, \hat{\alpha}_{\tiny \mbox{CK}}-\dot{\hat{\alpha}}_{\tiny \mbox{CK}} \, \eta \right)^2e^{2\gamma t}+\left( \frac{\eta}{\hat{\alpha}_{\tiny \mbox{CK}}} \right)^2 \right]\, .
\end{equation}
There are, however, serious points of criticism raised against this quantized version of the approach. Particularly, the uncertainty product of position and {\it physical} momentum seems to violate Heisenberg's principle, as it is decaying exponentially. On the other hand, Yu and Sun~\cite{Yu-Sun} showed that the Caldirola--Kanai (CK) Hamiltonian operator can be derived from the conventional system-plus-reservoir approach of Caldeira--Leggett. This puzzle will be solved below where the relation will be shown between the CK-approach and another effective model, using a (logarithmic) NLSE on the physical level. But previously, a different canonical approach, using an expanding coordinate system, shall be mentioned. 

One other point of criticism raised against the CK-approach is that $\hat{H}_{\tiny \mbox{CK}}$ is not a constant of motion. However, it can be turned into such an invariant by adding the term $\frac{\gamma}{2} x \, p \, e^{\gamma t}$ to it, leading to (in the following for an oscillator potential with constant frequency $\omega=\omega_0$)
\begin{equation}
\left[ \frac{1}{2m} p^2 + \frac{\gamma}{2} x p + \frac{m}{2}\omega_0^2x^2 \right]e^{\gamma t}=const\, ,
\end{equation}
if $x(t)$ obeys the equation of motion including the friction force. This invariant can be rewritten in a form like a conventional Hamiltonian, if a new (canonical) position coordinate and the corresponding (canonical) momentum are introduced via~\cite{elaf2010} 
\begin{eqnarray}
\hat{Q}&=&xe^{\frac{\gamma}{2}t}\, ,\nonumber \\
\hat{P}&=&m\dot{\hat{Q}}=m\left(\dot{x}+\frac{\gamma}{2}x\right)e^{\frac{\gamma}{2}t}\, .
\label{fis-exp}
\end{eqnarray}
The Hamiltonian then takes the form
\begin{equation}
\hat{H}_{\tiny \mbox{E}}=\frac{1}{2m} \hat{P}^2+\frac{m}{2}\left( \omega_0^2 - \frac{\gamma^2}{4}\right)\hat{Q}^2\, 
\label{H-exp}
\end{equation}
and is not only an invariant but, for $x_0=0$ or $p_0=m\dot{x}(0)=0$, even identical with the initial energy of the system. 

Hamiltonian (\ref{H-exp}) looks like that of an undamped HO with shifted frequency $\Omega^2=\omega_0^2-\frac{\gamma^2}{4}$ and the corresponding equation of motion for $\hat{Q}$ is consequently
\begin{equation}
\ddot{\hat{Q}}+\Omega^2\hat{Q}=0\, ,
\label{Newton-Exp}
\end{equation}
what, expressed in the physical coordinates, again provides the averaged Langevin equation (here for the oscillator)
\begin{equation}
\ddot{x}+\gamma\dot{x}+\omega_0^2x=0\, .
\end{equation}

Because we now have two canonical descriptions of the same physical dissipative system, the question arises, what is the connection between the CK~-~variables and the expanding ones, i.e., between $(\hat{x},\hat{p})$ and $(\hat{Q}=\hat{x}e^{\frac{\gamma}{2}t}, \hat{P}=\hat{p}e^{-\frac{\gamma}{2} t}+m\frac{\gamma}{2}\hat{x}e^{\frac{\gamma}{2}t})$? It is straightforward to show (e.g., via the Jacobian determinant being equal to one) that the transition between the two descriptions (on the canonical level) is given via a canonical transformation. Why is then $\hat{H}_{\tiny \mbox{E}}$ a constant of motion whereas $\hat{H}_{\tiny \mbox{CK}}$ is not? Because the generating function $\hat{F}_{2}(\hat{x},\hat{P}, t)$ leading to $\hat{H}_{\tiny \mbox{E}}$ via 
\begin{equation}
\hat{H}_{\tiny \mbox{E}}=\hat{H}_{\tiny \mbox{CK}}+\frac{\partial}{\partial t} \hat{F}_{2}
\end{equation}
is explicitly TD,
\begin{equation}
\hat{F}_{2}(\hat{x},\hat{P},t)=\hat{x}\hat{P}e^{\frac{\gamma}{2}t}-m\frac{\gamma}{4}\hat{x}^2e^{\gamma t}\, .
\end{equation}
The explicit time derivative $\frac{\partial}{\partial t}\hat{F}_{2}$ provides the missing contribution that turns $\hat{H}_{\tiny \mbox{CK}}$ into the invariant $\hat{H}_{\tiny \mbox{E}}$.

Consequently, also the action function $\hat{S}_{\tiny \mbox{E}}(\hat{Q},t)$ and $\hat{S}_{\tiny \mbox{CK}}(\hat{x},t)$ are different (what will be important in the quantum mechanical case considered below, due to the relation between the action function and the wave function, introduced by Schr\"odinger~\cite{Schrodinger}) and related via 
\begin{equation}
\hat{S}_{\tiny \mbox{CK}}(\hat{x},t)=\hat{S}_{\tiny \mbox{E}}(\hat{Q}(\hat{x},t),t)-m\frac{\gamma}{4}\hat{x}^2e^{\gamma t}\, . 
\label{actions-CK-E}
\end{equation}
As we are working on the canonical level, and the Hamiltonian (\ref{H-exp}) is free of constrains, Dirac quantization~\cite{D} can be achieved in the coordinates representation by keeping the position operator as a $c$-number, $\hat{Q}_{\tiny \mbox{op}}=\hat{Q}$, and expressing the momentum operator as $\hat{P}_{\tiny \mbox{op}}=\frac{\hbar}{i}\frac{\partial }{\partial \hat{Q}}$. The resulting canonical SE then has the form
\begin{equation}
i\hbar\frac{\partial}{\partial t}\hat{\Psi}_{\tiny \mbox{E}}=\left[-\frac{\hbar^2}{2m}\frac{\partial^2}{\partial \hat{Q}^2}+\frac{m}{2}\left( \omega_0^2 -\frac{\gamma^2}{4} \right)\hat{Q}^2\right]\hat{\Psi}_{\tiny \mbox{E}}\, .
\label{SE-E}
\end{equation}
The exact analytic solution in form of a Gaussian WP can be written as
\begin{equation}
\hat{\Psi}_{\tiny \mbox{E}}(\hat{Q},t)=\hat{N}_{\tiny \mbox{E}}(t)\exp\left\{ i\left[ \hat{y}_{\tiny \mbox{E}}(t)\tilde{Q}^2+\frac{1}{\hbar}\langle \hat{P} \rangle_{\tiny \mbox{E}}(t)\tilde{Q}+\hat{K}_{\tiny \mbox{E}}(t) \right] \right\}
\label{WP-E}
\end{equation}
with $\tilde{Q}=\hat{Q}-\langle \hat{Q} \rangle_{\tiny \mbox{E}}(t)$, $\langle \hat{P} \rangle_{\tiny \mbox{E}}(t)=m\dot{\langle \hat{Q} \rangle}_{\tiny \mbox{E}}(t)$ where $\langle ... \rangle_{\tiny \mbox{E}} $ now indicates that the mean values are calculated with the canonical wave function $\hat{\Psi}_{\tiny \mbox{E}}(\hat{Q},t)$, and $\hat{y}_{\tiny \mbox{E}}(t)$ is again a complex function of time.

The normalization factor $\hat{N}_{\tiny \mbox{E}}(t)$ and the phase factor $\hat{K}_{\tiny \mbox{E}}(t)$ are purely TD and again not are relevant for the equation of motion determining the evolution of the maximum and the width of the WP.

Inserting WP (\ref{WP-E}) into Eq. (\ref{SE-E}) provides the equation of motion for the maximum as
\begin{equation}
\frac{d^2}{dt^2}\langle \hat{Q} \rangle_{\tiny \mbox{E}}+\left( \omega_0^2-\frac{\gamma^2}{4} \right)\langle \hat{Q} \rangle_{\tiny \mbox{E}}=0\, ,
\label{New-Exp-eq}
\end{equation}
what, expressed in terms of the physical position variables $\langle x \rangle(t)=\eta(t)$ attains again the form of the Eq. (\ref{Dis-New-Eq}), i.e., a damped HO. 

The equation of motion for the WP width, depending of the complex variable $\hat{y}_{\tiny \mbox{E}}(t)$, can be expressed in terms of $\hat{C}_{\tiny \mbox{E}}=\frac{2\hbar}{m} \hat{y}_{\tiny \mbox{E}}(t)$ in the form of the Riccati equation  
\begin{equation}
\dot{\hat{C}}_{\tiny \mbox{E}}+\hat{C}_{\tiny \mbox{E}}^2+\left( \omega_0^2-\frac{\gamma^2}{4} \right)=0\, ,
\label{Riccati-Exp}
\end{equation}
where the imaginary part of $\hat{C}_{\tiny \mbox{E}}(t)$ is connected with the position uncertainty for the dissipative case on the canonical level in the same way as for the non-dissipative case on the physical level, i.e., $\hat{C}_{\tiny \mbox{E,I}}(t)=\frac{\hbar}{2m}\frac{1}{\langle \tilde{Q}^2 \rangle_{\tiny \mbox{E}}}$ with $\langle \tilde{Q}^2 \rangle_{\tiny \mbox{E}}~=~\langle \hat{Q}^2 \rangle_{\tiny \mbox{E}}-\langle \hat{Q} \rangle^2_{\tiny \mbox{E}}$. 

Equation (\ref{Riccati-Exp}) is identical with equation (4) of part I, only $C(t)$ is replaced by $\hat{C}_{\tiny \mbox{E}}(t)$ and $\omega^2$ by $\omega_0^2-\frac{\gamma^2}{4}$. Therefore, by introducing a new variable $\hat{\alpha}_{\tiny \mbox{E}}(t)$ via $\hat{C}_{\tiny \mbox{E,I}}(t)=\frac{1}{\hat{\alpha}^2_E(t)}$, in the same way as described in part I, the complex Riccati equation (\ref{Riccati-Exp}) can be transformed into the real NL Ermakov equation
\begin{equation}
\ddot{\hat{\alpha}}_{\tiny \mbox{E}}+\left( \omega_0^2-\frac{\gamma^2}{4}\right)\hat{\alpha}_{\tiny \mbox{E}}=\frac{1}{\hat{\alpha}_{\tiny \mbox{E}}^3}\, .
\end{equation}
Following the procedure outlined in part I, via the elimination of $\omega_0^2-\frac{\gamma^2}{4}$ between this equation and Eq.~(\ref{New-Exp-eq}), one obtains a dynamical Ermakov invariant in the form
\begin{equation}
\hat{I}_{\tiny \mbox{E}}=\frac{m}{2\hbar}\left[ \left(\dot{\langle \hat{Q} \rangle}_{\tiny \mbox{E}}\hat{\alpha}_{\tiny \mbox{E}}-\langle \hat{Q} \rangle_{\tiny \mbox{E}}\dot{\hat{\alpha}}_{\tiny \mbox{E}} \right)^2+\left( \frac{\langle \hat{Q} \rangle_{\tiny \mbox{E}}}{\hat{\alpha}_{\tiny \mbox{E}}} \right)^2 \right]\, .
\end{equation}
Expressing $\langle \hat{Q} \rangle_{\tiny \mbox{E}}(t)$ and $\dot{\langle \hat{Q} \rangle}_{\tiny \mbox{E}}(t)$, by means of Eq.~(\ref{fis-exp}), in terms of the physical variables $\langle x \rangle(t)=\eta(t)$ and $\dot{\langle x \rangle}(t)=\dot{\eta}(t)$, the Ermakov invariant can be rewritten as 
\begin{equation}
\hat{I}_{\tiny \mbox{E}}=\frac{m}{2\hbar} \, e^{\gamma t} \left[ \left(\dot{\eta}\hat{\alpha}_{\tiny \mbox{E}}-\left[\dot{\hat{\alpha}}_{\tiny \mbox{E}}-\frac{\gamma}{2}\hat{\alpha}_{\tiny \mbox{E}}\right]\eta \right)^2+\left( \frac{\eta}{\hat{\alpha}_{\tiny \mbox{E}}} \right)^2 \right]\, .
\label{Inv-Exp}
\end{equation}
How are these results related with the ones of the CK-approach? To answer this question, Schr\"odinger's original definition\footnote{In the first definition, the factor $i=\sqrt{-1}$ was missing, as Schr\"odinger considered $\Psi(x,t)$ to be real. The purely imaginary number was introduced in the third paper~\cite{E} of the Schr\"odinger's celebrated series on ``quantization as a problem of proper values''~\cite{F}.} of the wave function $\Psi(x,t)$ in terms of the action $S(x,t)$ (see~\cite{Schrodinger, F}), 
\begin{equation}
S(x,t)=\frac{\hbar}{i}\ln\Psi(x,t)\, ,
\end{equation}
can be used. According to this definition, Eq. (\ref{actions-CK-E}) can be translated into
\begin{equation}
\hat{\Psi}_{\tiny \mbox{CK}}(\hat{x},t)=\exp\left\{ -\frac{im}{\hbar}\frac{\gamma}{4}\hat{x}^2e^{\gamma t}\right\}\hat{\Psi}_{\tiny \mbox{E}}(\hat{Q}(\hat{x},t),t)\, ,
\end{equation}
thus defining the unitary transformation between the two approaches. 

This transformation requires that the complex quantity $\hat{C}_{\tiny \mbox{CK}}(t)$ determining the evolution of the WP width in the CK-system has to be connected with $\hat{C}_{\tiny \mbox{E}}(t)$ fulfilling Eq. (\ref{Riccati-Exp}) in the expanding system via 
\begin{equation}
\hat{C}_{\tiny \mbox{CK}}(t)=\left( \hat{C}_{\tiny \mbox{E}}(t)-\frac{\gamma}{2} \right)e^{\gamma t}\, .
\end{equation}
Inserting this into Eq. (\ref{Riccati-Exp}) actually turns it into the Riccati equation (\ref{Riccati-CK}) of the CK-approach. 


\subsection{ Non-canonical (nonlinear) description of dissipative systems}
\label{section-2.2}

From the previous discussion one concludes that the connection of the formal canonical description with the one on the physical level is still missing. To achieve this connection, the Brownian motion scenario is now considered not from the trajectory point of view, involving the (averaged) Langevin equation, but from the point of view of the probability distribution function, here in a quantum mechanical context, involving the inclusion of an additional diffusion process. In particular in position space this leads, as mentioned above, to the Smoluchowski equation. This changes the quantum mechanical continuity equation for $\rho_{\tiny \mbox{NL}}(x,t)=\Psi_{\tiny \mbox{NL}}(x,t)\Psi_{\tiny \mbox{NL}}^\ast(x,t)$ into
\begin{equation}
\frac{\partial \rho_{\tiny \mbox{NL}}}{\partial t}+\frac{\partial}{\partial x}\left[\rho_{\tiny \mbox{NL}} \, v_{\tiny \mbox{NL}}\right]-D_x\frac{\partial^2 \rho_{\tiny \mbox{NL}}}{\partial x^2}=0 
\label{Flokker-Planck-equation}
\end{equation}
with the (possibly TD) diffusion coefficient $D_x$, and the velocity field 
\begin{equation}
v_{\tiny \mbox{NL}}(x,t)=\frac{\hbar}{2im} \left[ \frac{1}{\Psi_{\tiny \mbox{NL}}(x,t)}\frac{\partial}{\partial x}\Psi_{\tiny \mbox{NL}}(x,t)-\frac{1}{\Psi_{\tiny \mbox{NL}}^\ast(x,t)}\frac{\partial}{\partial x}\Psi_{\tiny \mbox{NL}}^\ast(x,t) \right]\, .
\label{vel-field}
\end{equation}
Following the idea of Madelung~\cite{Madelung} and Mrowka~\cite{Mrowka} who obtained the complex Schr\"odinger equation via separation of the continuity equation into two complex conjugate equations, the same is tried with the Smoluchowski equation. Due to a coupling of $\Psi_{\tiny \mbox{NL}}(x,t)$ and $\Psi_{\tiny \mbox{NL}}^\ast(x,t)$ terms via the diffusion term, this separation is now not possible in general, but for particular cases. One of these is represented by the separation ansatz
\begin{equation}
-D_x \frac{\frac{\partial^2}{\partial x^2}\rho_{\tiny \mbox{NL}}}{\rho_{\tiny \mbox{NL}}}=\gamma \left[ \ln{\rho_{\tiny \mbox{NL}}}-\langle \ln{\rho_{\tiny \mbox{NL}}} \rangle \right]\, .
\label{Difusion-term}
\end{equation}
In this case the separation is possible, leading to an additional complex logarithmic term in the SE and therefore  to the NLSE
\begin{equation}
i\hbar\frac{\partial}{\partial t}\Psi_{\tiny \mbox{NL}}=\left[ -\frac{\hbar^2}{2m}\frac{\partial^2}{\partial x^2}+V(x)+\gamma\frac{\hbar}{i}\left[ \ln\Psi_{\tiny \mbox{NL}}-\langle \ln\Psi_{\tiny \mbox{NL}} \rangle_{\tiny \mbox{NL}} \right] \right]\Psi_{\tiny \mbox{NL}}\, .
\label{NLSE}
\end{equation}
This NLSE also possesses exact Gaussian WP solutions where the maximum follows the classical trajectory determined by the averaged Newtonian equation of motion (\ref{Dis-New-Eq}) for $\langle x \rangle_{\tiny \mbox{NL}}(t)=\eta(t)$.

The additional friction force in (\ref{Dis-New-Eq}) can be traced back to the real part of the logarithmic nonlinearity (which is identical with a friction term proposed by Kostin \cite{Kostin}, that, by itself, suffered from several shortcomings which are removed by the imaginary part of our NL term).

The WP width obeys again a complex Riccati equation, but now with an additional linear term proportional to the friction coefficient $\gamma$,
\begin{equation}
\dot{C}_{\tiny \mbox{NL}}+\gamma\,C_{\tiny \mbox{NL}}+C^2_{\tiny \mbox{NL}}+\omega^2=0\, .
\label{Riccati-NL}
\end{equation}
With the same definition of the variable $\alpha_{\tiny \mbox{NL}}(t)$ as in the non-dissipative case, i.e., $\hat{C}_{\tiny \mbox{NL,I}}=\frac{1}{\alpha^2_{\tiny \mbox{NL}}(t)}$, Eq. (\ref{Riccati-NL}) can be transformed into the Ermakov equation 
\begin{equation}
\ddot{\alpha}_{\tiny \mbox{NL}}+\left( \omega^2 -\frac{\gamma^2}{4} \right)\alpha_{\tiny \mbox{NL}}=\frac{1}{\alpha_{\tiny \mbox{NL}}^3}\, ,
\label{Damped-Ermakov-equation}
\end{equation}
which is formally identical with the one in the expanding coordinate system on the canonical level. Therefore, it is not surprising that the combination of this equation with the Newtonian equation (\ref{Dis-New-Eq}) leads exactly to the Ermakov invariant as written in (\ref{Inv-Exp}), only $\hat{\alpha}_{\tiny \mbox{E}}(t)$ has to be replaced by $\alpha_{\tiny \mbox{NL}}(t)$, both fulfilling the same differential equation.

The link between this physical level and the aforementioned canonical one can be found by applying again Schr\"odinger's definition of the action in terms of the wave function $\Psi(x,t)$~\cite{Schrodinger,F}. Starting point for Schr\"odinger's derivation of his equation was the Hamilton--Jacobi equation, here written in the TD form as
\begin{equation}
\frac{\partial}{\partial t}S+H\left(x, \frac{\partial}{\partial x}S, t \right)=0
\end{equation}
with the action function $S(x,t)$ and the momentum $p=\frac{\partial}{\partial x}S$, where we are now dealing with complex quantities. With his definition $S(x,t)=\frac{\hbar}{i}\ln\Psi(x,t)$ and a variational ansatz, he finally arrived at the Hamiltonian operator
\begin{equation}
H_{\tiny \mbox{L}}=-\frac{\hbar^2}{2m}\frac{\partial^2}{\partial x^2}+V(x)\, .
\end{equation}

Starting now from the NLSE (\ref{NLSE}) and reversing Schr\"odinger's procedure, one arrives at
\begin{equation} 
\left[ \frac{\partial}{\partial t}+\gamma \right]S + H_{\tiny \mbox{L}}\left(x, \frac{\partial}{\partial x}S, t \right)=\gamma\langle S \rangle\, .
\label{HJ-Eq}
\end{equation}
This is, of course, as little rigorous as Schr\"odinger's first attempt was. However, it follows his idea of connecting the classical Hamilton--Jacobi theory with a wave (mechanical) equation. The purely TD term $\gamma \langle S \rangle$ is necessary mainly for normalization purpose (can therefore be absorbed by the normalisation coefficient) and is neglected in the following. 

Multiplying the remaining Eq. (\ref{HJ-Eq}) by $e^{\gamma t}$ and using the definitions
\begin{equation}
\hat{S} = e^{\gamma t}S, \quad \hat{H}=e^{\gamma t}H_{{\tiny \mbox{L}}}\, ,
\end{equation}
it can be written as canonical Hamilton--Jacobi equation
\begin{equation}
\frac{\partial}{\partial t} \hat{S}+\hat{H}=0\, .
\end{equation}
From the definition of the action function it follows that the wave function $\hat{\Psi}(x,t)$ on the  canonical level is connected with the wave function $\Psi_{\tiny \mbox{NL}}(x,t)$ on the physical level via the \emph{non-unitary} relation
\begin{equation}
\ln\hat{\Psi}(x,t)=e^{\gamma t} \ln \Psi_{\tiny \mbox{NL}}(x,t)\, .
\label{NonU-Tras}
\end{equation}
Consequently, the momenta in the two systems are connected via 
\begin{equation}
\hat{p} = \frac{\hbar}{i} \frac{\partial}{\partial x}\ln\hat{\Psi}(x,t) = e^{\gamma t} \frac{\hbar}{i} \frac{\partial}{\partial x}\ln\Psi_{\tiny \mbox{NL}}(x,t)=e^{\gamma t} p\, ,
\end{equation}
which is equivalent to the connection between the canonical and the physical momentum in the CK-approach. 

The \emph{non-canonical} connection between the classical variables $(x,p)$ and the canonical ones $(\hat{x}=x,\hat{p}=e^{\gamma t}p)$ correspond to the \emph{non-unitary} transformation between $\Psi_{\tiny \mbox{NL}}(x,t)$ and $\hat{\Psi}_{\tiny \mbox{CK}}(x,t)$\footnote{Although $\Psi_{\tiny \mbox{NL}}(x,t)$ and $\Psi_{\tiny \mbox{CK}}(x,t)$ depend explicitly on the same variables, $x$ and $t$, the two wave functions are analytically different functions of $x$ and $t$ and have different physical meanings due to the non-unitary transformation (\ref{NonU-Tras}). Keeping this in mind, the apparent violation of the uncertainty relation mentioned above in the CK-approach does not occur any more (for details see~\cite{elaf2010}).}.

Expressing $\hat{H}$ in terms of the canonical momentum, in the classical case this leads to the CK-Hamiltonian (\ref{H-CK}) and in the quantum mechanical case to the modified SE (\ref{SE-CK}).

The non-unitary transformation (\ref{NonU-Tras}) between the wave function $\Psi_{\tiny \mbox{NL}}(x,t)$ on the physical level and $\hat{\Psi}_{\tiny \mbox{CK}}(x,t)$ on the canonical level also shows that the quantities fulfilling the complex Riccati equations for the WP width, i.e., Eq. (\ref{Riccati-NL}) for $C_{\tiny \mbox{NL}}(t)$ and Eq. (\ref{Riccati-CK}) for $\hat{C}_{\tiny \mbox{CK}}(t)$, are connected via 
\begin{equation}
\hat{C}_{\tiny \mbox{CK}}(t)=C_{\tiny \mbox{NL}}(t)e^{\gamma t}
\end{equation}
or, the corresponding Ermakov variables via 
\begin{equation}
\hat{\alpha}_{\tiny \mbox{CK}}(t)=\alpha_{\tiny \mbox{NL}}(t)e^{-\frac{\gamma}{2} t}\, .
\end{equation}

So, consequently, the equations for the mean value and uncertainties are uniquely related with  the corresponding equations on the physical level, obtained from the NLSE (\ref{NLSE}) and also connected with the system-plus-reservoir approach via the relation between the Caldeira--Leggett model and the one of the CK, as shown by Sun and Yu~\cite{Yu-Sun}. Furthermore, the equation of motion for the maximum and width of the WP describing the dissipative system and obtained from the NLSE (\ref{NLSE}) are also identical with the corresponding equations obtained from a different (but related) NLSE of Hasse~\cite{Hasse} (for details see~\cite{elaf2010}).

Therefore, the two equations that are relevant for the discussion of the effective description of the dissipative quantum system are the Newtonian equation
\begin{equation}
\ddot{\eta}+\gamma\,\dot{\eta}+\omega^2\eta=0
\label{N-E}
\end{equation}
and the complex Riccati equation
\begin{equation}
\dot{C}_{\tiny \mbox{NL}}+\gamma\,C_{\tiny \mbox{NL}}+C^2_{\tiny \mbox{NL}}+\omega^2 =0\, ,
\label{Ricc-NL}
\end{equation}
or its transformed Ermakov version 
\begin{equation}
\ddot{\alpha}_{\tiny \mbox{NL}}+\left( \omega^2-\frac{\gamma^2}{4} \right)\alpha_{\tiny \mbox{NL}}=\frac{1}{\alpha_{\tiny \mbox{NL}}^3}\, .
\end{equation}
These equations are discussed in detail subsequently.

To exploit formal similarities with the non-dissipative case, also the version of Eq. (\ref{Dis-New-Eq}) in expanding coordinates, corresponding to Eq. (\ref{Newton-Exp}), will be applied.


\section{Time-dependence of the uncertainties in the dissipative case}

\label{section-3}

Like in the non-dissipative case, the Riccati equation (\ref{Riccati-NL}) can be solved directly via transformation into a Bernoulli equation using the ansatz\footnote{As in this section essentially quantities connected with the NLSE (\ref{NLSE}) are discussed, the subscript ``NL'' of, e.g., $C_{\tiny \mbox{NL}}$, $\alpha_{\tiny \mbox{NL}}$, $\langle \cdots \rangle_{\tiny \mbox{NL}}$, etc, will be dropped.} 
\begin{equation}
C(t)=\tilde{C}+V(t)
\end{equation}
yielding
\begin{equation}
\dot{V}+2 \left(\tilde{C}+\frac{\gamma}{2}\right)V+V^2=0\, .
\label{Bern-eq}
\end{equation}
This equation can be linearized via the transformation $V(t)=\kappa(t)^{-1}$ to
\begin{equation}
\dot{\kappa}-2\left(\tilde{C}+\frac{\gamma}{2}\right)\kappa=1\, .
\end{equation}
For the harmonic potential with constant frequency $\omega_0$ the particular solution $\tilde{C}$ is constant and has the form
\begin{equation}
\tilde{C}_{\pm}=-\frac{\gamma}{2}\pm\sqrt{\frac{\gamma^2}{4}-\omega_0^2}\, .
\end{equation}
Note, that (unlike in the non-dissipative case) also for $\omega_0=0$ (i.e., the damped free motion) two different particular solutions exist. The solutions of the Riccati equation can then be expressed as $C_{\pm}(t)=\tilde{C}_{\pm}+V_{\pm}(t)$ with
\begin{equation}
V_{\pm}(t)=\frac{1}{\kappa_{\pm}(t)}=\frac{e^{-2(\tilde{C}_{\pm}+\frac{\gamma}{2})t}}{\kappa_0+\frac{1}{2(\tilde{C}_{\pm}+\frac{\gamma}{2})}\left[ 1-e^{-2(\tilde{C}_{\pm}+\frac{\gamma}{2})t}\right]}
\label{Bern-int-cond}
\end{equation}
with $\tilde{C}_{\pm}+\frac{\gamma}{2}=\pm\sqrt{\frac{\gamma^2}{4}-\omega^2_0}$\, .

As in the non-dissipative case, the choice of the initial conditions (where $\kappa_0$ again depends on the initial uncertainties and the initial correlation, or $\alpha(t_0)=\alpha_0$ and $\dot{\alpha}(t_0)=\pm|\dot{\alpha}_0|$, respectively) can strongly influence the dynamics of the quantum uncertainties and the physical properties depending thereon (examples are shown below, in Section \ref{section-6}).

In order to find the relations between $\kappa_0$ and the initial properties of the WP, one can use the solution $C(t)$ of the complex Riccati equation (\ref{Riccati-NL}), expressed in terms of $\alpha(t)$ that fulfills Eq. (\ref
{Damped-Ermakov-equation}), i.e.
\begin{equation}
C(t) = \frac{\dot{\alpha}(t)}{\alpha(t)}-\frac{\gamma}{2}+i\frac{1}{\alpha^2(t)}=C_{\tiny \mbox{R}}(t)+i C_{\tiny \mbox{I}}(t)\, .
\label{Riccati-solution}
\end{equation}
The uncertainties can then be written as
\begin{eqnarray}
\sigma^2_x(t) &=&\frac{\hbar}{2m} \alpha^2(t)=\frac{\hbar}{2m}\, \frac{1}{C_{\tiny \mbox{I}}(t)}\, , \label{damping-sigmax}\\ 
\sigma^2_p(t)&=&\frac{m \hbar}{2} \left[ \left( \dot{\alpha}(t)-\frac{\gamma}{2}\alpha(t) \right)^2+\frac{1}{\alpha^2(t)} \right]=\frac{m\hbar}{2}\, \frac{C_{\tiny \mbox{R}}^2(t)+C_{\tiny \mbox{I}}^2(t)}{C_{\tiny \mbox{I}}(t)}\, , \label{damping-sigmap}  \\
\sigma_{xp}(t)&=&\frac{\hbar}{2}\alpha(t)\left[ \dot{\alpha}(t)-\frac{\gamma}{2}\alpha(t) \right]= \frac{\hbar}{2}\, \frac{C_{\tiny \mbox{R}}(t)}{C_{\tiny \mbox{I}}(t)}\, .
\label{damping-sigmaxp}
\end{eqnarray} 
It is straightforward to prove that these expressions minimize the Schr\"odinger-Robertson uncertainty relation
\begin{equation}
\sigma_x^2(t)\sigma_p^2(t)-\sigma_{xp}^2(t)=\frac{\hbar^2}{4}\, ,
\end{equation}
as in the non-dissipative case.

Crucial point is, like in the non-dissipative case, the determination of $\alpha(t)$ and $\dot{\alpha}(t)$ for a given initial condition $\alpha_0$ and $|\dot{\alpha}_0|$. For this purpose, the method of linear invariant operators used already in part I (see there Appendix B) is applied because the dissipative system can be formulated like a non-dissipative system  (only with shifted frequency) in the expanding coordinates framework.

Expressing $\eta(t)$ in the expanding system as $\xi(t)=\eta(t)e^{\frac{\gamma}{2}t}$ with the equation of motion 
\begin{equation}
\ddot{\xi}+\left(\omega_0^2-\frac{\gamma^2}{4} \right)\xi=0,
\label{Phy-New-Eq}
\end{equation}
the Ermakov invariant (\ref{Inv-Exp}) can be rewritten on the canonical level as
\begin{equation}
I=\frac{m}{2\hbar}\left[ \left( \dot{\xi}\alpha-\xi\dot{\alpha} \right)^2+\left(\frac{\xi}{\alpha} \right)^2 \right]\, .
\label{Inv-Phy}
\end{equation}
Following the procedure outlined in Appendix B of part I, i.e., equating the operator corresponding to (\ref{Inv-Phy}) with the most general quadratic invariant operator (Eq. (B.5) part I) finally leads to the expression for $\alpha(t)$ in the form
\begin{equation}
\alpha_{\mp}(t)=\sqrt{m\hbar}\left[ A\xi_1^2(t)+B\xi_2^2(t)\mp2C\xi_1(t)\xi_2(t) \right]^{\frac{1}{2}}\, ,
\label{Erm-New}
\end{equation}
where $\xi_1(t)$ and $\xi_2(t)$ are two linear independent solutions of Eq. (\ref{Phy-New-Eq}). The constants $A$, $B$ and $C$, expressed in terms of $\alpha_0$ and $\vert\dot{\alpha}_0\vert$, have the same form as in the non-dissipative case, i.e.,
\begin{equation}
A=\frac{m}{\hbar}\left(|\dot{\alpha}_0|^2+\frac{1}{\alpha_0^2} \right)\, , \quad B=\frac{\alpha_0^2}{\hbar m}\, , \quad C=\frac{|\dot{\alpha}_0|\alpha_0}{\hbar}\, .
\label{const-alpha}
\end{equation}
The double sign $\mp$ in Eq. (\ref{Erm-New}) implies different initial conditions for $\dot{\alpha}(t)$, the negative sign corresponds to the initial condition $\dot{\alpha}(t_0)=|\dot{\alpha}_0|$ whereas the positive sign to the initial condition $\dot{\alpha}(t_0)=-|\dot{\alpha}_0|$. Unlike in non-dissipative systems, the choice of the sign is fundamental in the evolution of the WPs, as is proved below.

The initial values for $\xi_i(t)$ are given by
\begin{equation}
\xi_1(t_0)=0\, ,\quad\dot{\xi}_1(t_0)=-\frac{1}{m}\, ,\quad\xi_2(t_0)=1\, ,\quad\dot{\xi}_2(t_0)=0\, .
\label{init-cond-New}
\end{equation}
From~(\ref{Riccati-solution}), the initial value for the Riccati variable, $C_0$, can be expressed in terms of $\alpha_0$ and $|\dot{\alpha}_0|$ or in terms of the uncertainties as
\begin{equation}
C_0=\mp\frac{|\dot{\alpha}_0|}{\alpha_0}-\frac{\gamma}{2}+\frac{i}{\alpha_0^2}=\frac{1}{m}\frac{\sigma_{xp_0}}{\sigma_{x_0}^2}+\frac{i\hbar}{2m}\frac{1}{\sigma^2_{x_0}}=\frac{1}{m\sigma_{x_0}^2}\left( \sigma_{xp_0}+i\frac{\hbar}{2} \right)\, .
\end{equation}
Therefore, the initial condition entering the solution (\ref{Bern-int-cond})  of the Bernoulli equation (\ref{Bern-eq}) can be written in the form
\begin{eqnarray}
V_0=\frac{1}{\kappa_0}&=&C_0-\tilde{C}\nonumber\\
&=&\left( \mp\frac{|\dot{\alpha}_0|}{\alpha_0}-\frac{\gamma}{2} \right)-\left( -\frac{\gamma}{2}\pm \sqrt{\frac{\gamma^2}{4}-\omega_0^2} \right)+\frac{i}{\alpha_0^2}\nonumber\\
&=& \mp\frac{|\dot{\alpha}_0|}{\alpha_0}+i\left[ \frac{1}{\alpha_0^2}\mp\sqrt{\omega_0^2-\frac{\gamma^2}{4}} \right]\, .
\end{eqnarray}
Depending on $\frac{\gamma^2}{4}$ being larger, smaller or equal to $\omega_0$, the square root contributes to the real or imaginary part of $V_0$. 

The expectation value of the NL Hamiltonian, which is associated with the expression~(\ref{NLSE}), has the same functional form as in the non-dissipative case, i.e.,
\begin{equation}
E(t)=\frac{1}{2m}\langle p \rangle^2(t)+\frac{1}{2}m\omega^2\langle x \rangle^2(t)+\frac{1}{2m}\sigma_p^2(t)+\frac{m}{2}\omega_0^2\sigma_x^2(t)\, .
\end{equation} 
However the explicit expressions of the mean values and uncertainties are different. One can define the quantum contribution, $\tilde{E}(t)$, in terms of the solution of the Ermakov equation   
\begin{eqnarray}
 \tilde{E}(t)&=&\frac{1}{2m}\sigma_p^2(t)+\frac{m}{2}\omega_0^2\sigma_x^2(t) \,  \nonumber\\
 &=&\frac{ \hbar}{4}\left[ \left( \dot{\alpha}(t)-\frac{\gamma}{2}\alpha(t) \right)^2+\frac{1}{\alpha^2(t)}+\omega_0^2\alpha^2(t) \right] \, ,
 \label{energy}
\end{eqnarray}
which depends on the initial conditions $\alpha_0$ and $|\dot{\alpha}_0|$ or equivalently on the initial uncertainties.  Therefore the ambiguity of signs mentioned above leads to different dynamical properties unlike in the non-dissipative case. Examples are shown in Section~\ref{section-6}.


\section{Riccati--Newton connections and the propagator}

\label{section-4}

The complex Riccati equation (\ref{Riccati-NL}) can be linearized using the logarithmic derivative,
\begin{equation}
C_{\tiny \mbox{NL}} (t)=\frac{2\hbar}{m}y_{\tiny \mbox{NL}}(t)=\frac{\dot{\tilde{\lambda}}(t)}{\tilde{\lambda}(t)}\, ,
\end{equation}
which, expressed in terms of $\tilde{\lambda}(t)$, takes the form of the corresponding (complex) Newtonian equation (now including the friction term)
\begin{equation}
\ddot{\tilde{\lambda}}+\gamma\dot{\tilde{\lambda}}+\omega^2 \tilde{\lambda}=0\, .
\end{equation}

In polar coordinates $\tilde{\lambda}(t)$ can be written in the form $\tilde{\lambda}(t)=\alpha_{\tiny \mbox{NL}}(t)\,e^{-\frac{\gamma}{2}t+i\phi_{\tiny \mbox{NL}}(t)}$, thus the Riccati solution reads 
\begin{equation}
C_{\tiny \mbox{NL}}(t)=\frac{\dot{\alpha}_{\tiny \mbox{NL}}(t)}{\alpha_{\tiny \mbox{NL}}(t)}-\frac{\gamma}{2}+i\dot{\phi}_{\tiny \mbox{NL}}(t) \, .
\end{equation}
Then, the imaginary part of $C_{\tiny \mbox{NL}}(t)$ is given by $C_{\tiny \mbox{NL,I}}(t)=\dot{\phi}_{\tiny \mbox{NL}}(t)=\frac{1}{\alpha_{\tiny \mbox{NL}}^2(t)}$, i.e., the relation between $\dot{\phi}_{\tiny \mbox{NL}}(t)$ and $\alpha_{\tiny \mbox{NL}}^2(t)$ remains the same as in the non-dissipative case, see part I. 

We define an exponentially expanding variable $\lambda(t)=\tilde{\lambda}(t)e^{\frac{\gamma}{2}t}$ (like $\hat{Q}(t)$ compared to $x(t)$ or $\xi(t)$ compared to $\eta(t)$). Then the Riccati solution takes the form $C_{\tiny \mbox{NL}}(t)=\frac{\dot{\lambda}(t)}{\lambda(t)}-\frac{\gamma}{2}$ and the corresponding complex linear equation for $\lambda(t)$ is
\begin{equation}
\ddot{\lambda}+\left(\omega^2-\frac{\gamma^2}{4} \right)\lambda=0\, ,
\end{equation}
i.e., again like an undamped oscillator with shifted frequency. The polar form of this variable has the same form as in the non-dissipative case, $\lambda(t)=\alpha_{\tiny \mbox{NL}}(t)e^{i\phi_{\tiny \mbox{NL}}(t)}$; therefore, the conservation law: 
\begin{equation}
\dot{\phi}_{\tiny \mbox{NL}}(t)\alpha_{\tiny \mbox{NL}}^2(t)=\dot{\lambda}_{\tiny \mbox{I}}(t)\lambda_{\tiny \mbox{R}}(t)-\dot{\lambda}_{\tiny \mbox{R}}(t)\lambda_{\tiny \mbox{I}}(t)=1
\end{equation}
is also fulfilled. Additionally, following the procedure presented in part I, the real and the imaginary parts of $\tilde{\lambda}(t)$ satisfy
\begin{equation}
\left(
   \begin{array}{c}
      	\tilde{\lambda}_{\tiny \mbox{R}}(t)  \\
      \tilde{\lambda}_{\tiny \mbox{I}} (t) \\
   \end{array}
   \right)
   = 
	\left(
	 \begin{array}{cc} 
	      \mp c \, \alpha_{\tiny \mbox{NL}}(t) \, \dot{\alpha}_{\tiny \mbox{NL}}(t) & \pm \frac{c}{m} \alpha_{\tiny \mbox{NL}}^2(t) \\
	      c & 0 \\
	   \end{array}
	   \right)
              	     \left(
	   	             \begin{array}{c}
			     \langle x\rangle_{\tiny \mbox{NL}}(t)\\
		              \langle p\rangle_{\tiny \mbox{NL}}(t) + \frac{\gamma m}{2} \langle x\rangle_{\tiny \mbox{NL}}(t) \\
		               \end{array} 
                      \right)\, ,
                      \label{L-QM}
\end{equation}
where now the relations (\ref{damping-sigmax}) and (\ref{damping-sigmaxp}) are used together with the constant $c=\sqrt{\frac{m}{2\hbar I_{\tiny \mbox{NL}}}}$, with $I_{\tiny \mbox{NL}}$ given by (\ref{Inv-Phy}). Thus we have expressed the solutions of the complex Newtonian equation for dissipative systems in terms of quantum mechanical quantities, which also can be used to establish the initial conditions for the WP. 

The real and the imaginary parts of $\tilde{\lambda}(t)$ can, like in the non-dissipative case, be used to construct the TD Green function or Feynman kernel of the system by substituting them into Eq. (33) of part I, for further details see~\cite{CSCR}.  This result allows to study the evolution of any initial state in the dissipative system.

The Ermakov invariant, in terms of the real and imaginary parts of $\tilde{\lambda}(t)$, takes the form
\begin{equation}
I_{\tiny \mbox{NL}}=\frac{m}{2\hbar}\frac{e^{\gamma t}}{c^2}\left[ \left( \frac{\tilde{\lambda}_{\tiny \mbox{R}}}{\alpha_{\tiny \mbox{NL}}} \right)^2+\left( \frac{\tilde{\lambda}_{\tiny \mbox{I}}}{\alpha_{\tiny \mbox{NL}}} \right)^2 \right]\, ,
\end{equation}
which, using the relation~(\ref{L-QM}) can also be written as
\begin{equation}
I_{\tiny \mbox{NL}}=\frac{e^{\gamma t}}{\hbar^2}\left[ \sigma^2_p(t)\langle x\rangle_{\tiny \mbox{NL}}^2-2\sigma_{xp}(t)\langle x\rangle_{\tiny \mbox{NL}} \langle p \rangle_{\tiny \mbox{NL}}+\sigma^2_x(t)\langle p\rangle_{\tiny \mbox{NL}}^2 \right]\, .
\label{INV-NL}
\end{equation}
Thus the Ermakov invariant can be determined  in terms of the parameters of the WP.


\section{Wigner function}

\label{section-5}

In the classical description of dissipative systems through the Langevin stochastic differential equation
for phase space variables, it has been shown that a generic Fokker--Planck equation for the probability density function can be established~\cite{Chandrasekhar, Kerr}. We are going to show that the Wigner function associated to the dissipative WP solution of the NLSE satisfies a Fokker--Planck-type equation and the marginal in position space satisfies a Smoluchowski equation.

In quantum mechanics, the description of a state in phase space can be formulated by the Wigner function, defined by 
\begin{equation}
W(x,p;t)=\frac{1}{2\pi\hbar}\int_{-\infty}^\infty dq\,\Psi^\ast\left( x+\frac{q}{2},t \right)\Psi\left( x-\frac{q}{2},t \right)e^{\frac{i}{\hbar}q \, p}\, .
\end{equation}
Then, the Wigner function associated with $\Psi_{\tiny \mbox{NL}}(x,t)$ corresponds to 
\begin{equation}
W_{\tiny \mbox{NL}}(x,p;t)=\frac{1}{\pi\hbar}\exp\left\{ -\frac{2}{\hbar^2}\left[ \sigma_p^2(t)\tilde{x}^2 -2\sigma_{xp}(t)\tilde{x} \tilde{p} +\sigma_x^2(t)\tilde{ p }^2 \right] \right\} \, ,
\label{Ph-WF}
\end{equation}
where $\tilde{x}= x - \eta(t)$ and $\tilde{p}=p - m \dot{\eta}(t)$.
Therefore, the Wigner function for the dissipative case has the same functional form as in the non-dissipative case, see Eq. (37) part I; however, in this case the argument of the exponential does not correspond to the Ermakov invariant in the same form as in the non-dissipative case. We can express $W_{\tiny \mbox{NL}}(x,p;t)$ in terms of $I_{\tiny \mbox{NL}}$ recognising from expression (\ref{INV-NL}) that
\begin{equation}
  e^{-\gamma t} I_{\tiny \mbox{NL}}(\tilde{x},\tilde{p};t)=\frac{1}{\hbar^2}\left[ \sigma_p^2(t) \, \tilde{x}^2 -2\sigma_{xp}(t) \, \tilde{x} \, \tilde{p} +\sigma_x^2(t) \, \tilde{ p }^2 \right]\, ,
\end{equation}
wherefrom one obtains 
\begin{equation}
W_{\tiny \mbox{NL}}(x,p;t)=\frac{1}{\pi\hbar}\exp\left\{ -2  e^{-\gamma t} I_{\tiny \mbox{NL}}(\tilde{x},\tilde{p};t)\right\}.
\end{equation}

The partial differential equation that the Wigner function obeys in the dissipative case, as in the classical case, is a Fokker--Planck equation given by
{\small{
\begin{eqnarray}
\frac{\partial W_{\tiny \mbox{NL}}}{\partial t}+\frac{p}{m}\frac{\partial W_{\tiny \mbox{NL}}}{\partial x}-\left[ m\omega^2 \langle x \rangle_{\tiny \mbox{NL}}+\gamma\langle p \rangle_{\tiny \mbox{NL}} \right]\frac{\partial W_{\tiny \mbox{NL}}}{\partial p} 
-D_x \frac{\partial^2 W_{\tiny \mbox{NL}}}{\partial x^2}-D_p \frac{\partial^2 W_{\tiny \mbox{NL}}}{\partial p^2}=0\, 
\label{FPE-WF}
\end{eqnarray}
}}
where the diffusion coefficients are given by $D_x=\frac{\gamma}{2}\sigma_x^2(t)$ and $D_p=-\frac{\gamma}{2}\sigma_p^2(t)$, the mean value $\langle x \rangle_{\tiny \mbox{NL}}(t)$ satisfies the Langevin equation (\ref{N-E}), and the uncertainties satisfy the the set of ordinary differential equations~\cite{schu2}
\begin{eqnarray}
\frac{d \sigma_x^2}{dt}&=&\frac{2}{m}\sigma_{xp}+\gamma\,\sigma_x^2\, ,
\label{dispersion-1}\\
\frac{d \sigma_p^2}{dt}&=&-2\,m\,\omega^2\sigma_{xp}-\gamma\,\sigma_p^2\, ,
\label{dispersion-2}\\
\frac{d \sigma_{xp}}{dt}&=&\frac{1}{m}\sigma_p^2-m\,\omega^2\sigma_x^2 \, .
\label{dispersion-3}
\end{eqnarray} 
It is proved in \ref{Appendix-A} that the uncertainties and the correlation function defined by Eqs. (\ref{damping-sigmax}), (\ref{damping-sigmap}) and (\ref{damping-sigmaxp}) indeed satisfy the above closed set of differential equations. 

Furthermore, it is well-known in standard quantum theory that the mar\-gi\-nals of the Wigner function determine the density probabilities in position and in momentum representation, i.e.,  
\begin{equation}
\rho(x,t)= \int_{-\infty}^\infty dp\, W(x,p;t) \, ,\quad \tilde{\rho}(p,t) = \int_{-\infty}^\infty dx\, W(x,p;t)\, .
\end{equation} 
For the dissipative case this property of the Wigner function is still fulfilled. In order to prove this, the equation of motion for these distribution functions can be obtained from Eq. (\ref{FPE-WF}). So, integrating Eq. (\ref{FPE-WF}) with respect to the momentum $p$-variable, the equation for $\rho_{\tiny \mbox{NL}}(x,t)$ is attained,
\begin{equation}
\frac{\partial \rho_{\tiny \mbox{NL}}}{\partial t}+\frac{\partial}{\partial x}\left[\rho_{\tiny \mbox{NL}} \, v_{\tiny \mbox{NL}}\right]-D_x \frac{\partial^2 \rho_{\tiny \mbox{NL}}}{\partial x^2}=0\, ,
\end{equation}
where the velocity field is given by
\begin{eqnarray}
v_{\tiny \mbox{NL}}(x,t)&=&\frac{\hbar}{2im} \left[ \frac{1}{\Psi_{\tiny \mbox{NL}}(x,t)}\frac{\partial}{\partial x}\Psi_{\tiny \mbox{NL}}(x,t)-\frac{1}{\Psi_{\tiny \mbox{NL}}^\ast(x,t)}\frac{\partial}{\partial x}\Psi_{\tiny \mbox{NL}}^\ast(x,t) \right] \nonumber \\
&=&\frac{1}{m}\frac{\sigma_{xp}(t)}{\sigma_x^2(t)}\tilde{x}+\frac{d \langle x \rangle_{\tiny \mbox{NL}}}{dt}=C_{\tiny \mbox{NL,R}}(t) \, \tilde{x}+\frac{d \langle x \rangle_{\tiny \mbox{NL}}}{dt} \, ,
\end{eqnarray}
i.e, the Smoluchowski equation (\ref{Flokker-Planck-equation}) is recovered. 

On the other hand, integrating Eq. (\ref{FPE-WF}) with respect to the position $x$-variable one obtains 
\begin{equation}
\frac{\partial \tilde{\rho}_{\tiny \mbox{NL}}}{\partial t}+\frac{\partial}{\partial p}\left[\tilde{\rho}_{\tiny \mbox{NL}} \, \tilde{v}_{\tiny \mbox{NL}}\right]-D_p\frac{\partial^2\tilde{\rho}_{\tiny \mbox{NL}}}{\partial p^2}=0\, ,
\end{equation}
now the velocity field is given by
\begin{eqnarray}
\tilde{v}_{\tiny \mbox{NL}}(p,t)&=& \frac{\hbar m \omega^2}{2i}\left[ \frac{1}{\tilde{\Psi}_{\tiny \mbox{NL}}(p,t)}\frac{\partial}{\partial p}\tilde{\Psi}_{\tiny \mbox{NL}}(p,t)-\frac{1}{\tilde{\Psi}_{\tiny \mbox{NL}}^\ast(p,t)}\frac{\partial}{\partial p}\tilde{\Psi}_{\tiny \mbox{NL}}^\ast(p,t) \right]-\gamma\langle p \rangle_{\tiny \mbox{NL}} \nonumber \\
&=&-m\omega^2 \frac{\sigma_{xp}(t)}{\sigma_p^2(t)}\tilde{p}+\frac{d \langle p \rangle_{\tiny \mbox{NL}}}{dt}=-\omega^2\frac{C_{\tiny \mbox{NL,R}}(t)}{|C_{\tiny \mbox{NL}}(t)|^2}\tilde{p}+\frac{d \langle p \rangle_{\tiny \mbox{NL}}}{dt} \, , 
\end{eqnarray}
where $\tilde{\Psi}_{\tiny \mbox{NL}}(p,t)$  and $\Psi_{\tiny \mbox{NL}}(x,t)$ are related by a Fourier transformation. Therefore, this shows that truly the marginals of $W_{NL}(x,p;t)$ correspond to the probability distribution functions in position and momentum space.


\section{Examples}

\label{section-6}

In this section the sensitivity of the quantum mechanical properties of the dissipative system, like the uncertainties and the quantum contribution to the energy, to the choice of initial conditions is demonstrated. For this purpose, the damped free motion and the damped HO for the possible cases, under critical damping ($\omega_0 > \frac{\gamma}{2}$), overdamping ($\omega_0 < \frac{\gamma}{2}$) and aperiodic limit ($\omega_0=\frac{\gamma}{2}$) are considered.

For the determination of the aforementioned quantum mechanical properties the knowledge of $\alpha(t)$\footnote{As in Section~\ref{section-3}, in this section we are dealing essentially with quantities for the NLSE, thus the subscript ``NL'' will be dropped.}, i.e., the solution of the Ermakov equation (\ref{Damped-Ermakov-equation}), and its time-derivative $\dot{\alpha}(t)$ are needed. As mentioned in Section \ref{section-3} and shown explicitly in Appendix B of part I, $\alpha_(t)$ can be obtained from the most general quadratic invariant operator in the form (\ref{Erm-New}) that only requires two linear independent solutions $\xi_i(t)$ of the linear Newtonian equation (\ref{Phy-New-Eq}). In this way $\alpha^2(t)$ can be determined from $\xi_1(t)$ and $\xi_2(t)$. 

In principle, for $\alpha(t)$ being the square root, two signs, plus and minus, are possible. Considering $\alpha(t)$ being proportional to the WP width, only the positive sign is physical meaningful, but there is still a plus/minus sign in front of the term with coefficient $C$ under the square root, see Eq (\ref{Erm-New}), leading to possibly two different solutions for $\alpha(t)$. This ambiguity is not present for the initial condition $|\dot{\alpha}_0|=0$ because in this case the coefficient $C$ vanishes.

The general expressions of the uncertainties in term of $\xi_i(t)$ and its momenta $g_i(t)=-m \, \dot{\xi}_i(t)$ are obtained in \ref{Appendix-A} and shown in Table~\ref{Table-1}, where the initial conditions $\alpha_0$ and $|\dot{\alpha}_0|$ and the two possible signs for $|\dot{\alpha}_0|$ are taken into account.

\begin{table}[htp]
\begin{center}
\begin{tabular}{cc}
\hline
\\
$\sigma_x^2(t)$ & $\frac{\hbar}{2 m}\left[ m^2\beta_0\xi_1^2(t)+\left(\mp m|\dot{\alpha}_0|\xi_1(t)+\alpha_0\xi_2(t) \right)^2 \right]$ \\
\\
$\sigma_p^2(t)$ &$\frac{\hbar m}{2}\left[\chi_1^2(t)\beta_0+\left(\mp\chi_1(t)|\dot{\alpha}_0|+\chi_2(t)\frac{\alpha_0}{m} \right)^2\right]$\\
\\
$\sigma_{xp}(t)$ & $\frac{\hbar}{2}\left[ \left( m\left[|\dot{\alpha}_0|^2+\frac{1}{\alpha_0^2}\right]\mp|\dot{\alpha}_0|\alpha_0 \right)\chi_1(t)\xi_1(t)+\left(\frac{\alpha_0^2}{m}\mp|\dot{\alpha}_0|\alpha_0 \right)\chi_2(t)\xi_2(t)\right]$\\
\\
$\tilde{E}(t)$&$\frac{1}{2m}\sigma_p^2(t)+\frac{1}{2}m\omega^2_0\sigma_x^2(t)$\\
\\
$v_{tun}(x,t)$&$\frac{-\beta_0 \, \xi_1 \, g_1-\left(\mp|\dot{\alpha}_0| \, g_1(t)+\frac{\alpha_0}{m} \, g_2(t)\right)\left(\mp|\dot{\alpha}_0| \, \xi_1(t)+\frac{\alpha_0}{m} \, \xi_2(t)\right)}{\beta_0 \, \xi_1^2(t)+\left(\mp|\dot{\alpha}_0| \, \xi_1(t)+\frac{\alpha_0}{m} \, \xi_2(t)\right)^2} \, \frac{\tilde{x}}{m}$\\
\\
\hline
\end{tabular}
\end{center}
\caption{\footnotesize In this table the general expressions for the uncertainties, the correlation function, the energy and the tunnelling velocity in terms of the time-dependent functions $\xi_i(t)$ and its momenta $g_i(t)$ are given, where $\chi_i(t)=-g_i(t)-m\frac{\gamma}{2} \xi_i(t)$, with $i=1,2$.}
\label{Table-1}
\end{table}

In the non-dissipative case the selection of the sign was reflected by the sign of the correlation function and the direction of the tunnelling velocity. The change of sign in the dissipative case has more important implications due to the environmental parameter $\gamma$.  To enhance the differences between these cases we consider next the properties of the correlations, tunnelling currents, and the quantum contributions to the energy.

For the correlation function in the non-dissipative case the initial conditions $(\alpha_0,|\dot{\alpha}_0|)=(\alpha_0,0)$ corresponds to $\sigma_{xp_0}=\pm\frac{\hbar}{2}\dot{\alpha}_0|\alpha_0|=0$. In contrast, in the dissipative case, $\sigma_{xp_0}=\frac{\hbar}{2}\alpha_0\left( \pm|\dot{\alpha}_0|-\frac{\gamma}{2}\alpha_0 \right)=0$ requires the positive sign and $|\dot{\alpha}_0|=\frac{\gamma}{2}\alpha(t_0)\neq0$. The initial condition $|\dot{\alpha}_0|=0$ leads to a different expressions for the correlation of position and momentum, $\sigma_{xp_0}\neq0$, and therefore yields different physical properties. These choices of initial conditions are, of course, only two out of many, but they already demonstrate the influence of $\gamma$ on the physical properties of the quantum system.

The tunnelling current for the non-dissipative case was defined in the Appendix C of part I. There it was shown that the velocity field $v(x,t)$, corresponding to the convective probability current $j(x,t)=\rho(x,t)v(x,t)$ of the continuity equation, given by Eq. (\ref{vel-field}) of this paper and therefore depending on the phase of the Gaussian WP solution, $\Psi(x,t)$, was given by
\begin{equation}
v(x,t)=\dot{\eta}(t)+\frac{\dot{\alpha}_{\tiny \mbox{L}}(t)}{\alpha_{\tiny \mbox{L}}(t)}\tilde{x},
\end{equation}
where the second term $v_{\tiny \mbox{tun}}(x,t)=\frac{\dot{\alpha}_{\tiny \mbox{L}}(t)}{\alpha_{\tiny \mbox{L}}(t)}\tilde{x}$, is related with the tunnelling current $j_{\tiny \mbox{tun}}(x,t)=\rho(x,t)v_{\tiny \mbox{tun}}(x,t)$. 

In the dissipative case, the phase of $\Psi_{\tiny \mbox{NL}}(x,t)$ changes because $C_{\tiny \mbox{NL,R}}$ changes from $\frac{\dot{\alpha}_{\tiny \mbox{L}(t)}}{\alpha_{\tiny \mbox{L}}(t)}$ to $\frac{\dot{\alpha}_{\tiny \mbox{NL}}(t)}{\alpha_{\tiny \mbox{NL}}(t)}-\frac{\gamma}{2}$. Therefore, in this case the velocity field (\ref{vel-field}), denoted by $v_{NL}(x,t)$, attains the form
\begin{equation}
v_{\tiny \mbox{NL}}(x,t)=\left[\frac{\dot{\alpha}_{\tiny \mbox{NL}}(t)}{\alpha_{\tiny \mbox{NL}}(t)}-\frac{\gamma}{2}\right]\tilde{x} + \dot{\eta}(t).
\end{equation}
However, there is an additional contribution from the diffusion term 
\begin{equation}
-D_x\frac{\partial^2 \rho_{\tiny \mbox{NL}}}{\partial x^2}=\frac{\partial}{\partial x}\left[ -\rho_{\tiny \mbox{NL}}\left(\frac{D_x}{\rho_{\tiny \mbox{NL}}}\frac{\partial \rho_{\tiny \mbox{NL}}}{\partial x}\right) \right]
\end{equation} 
with a corresponding diffusion velocity 
\begin{equation}
v_{D_x}(x,t)=-\frac{D_x}{\rho_{\tiny \mbox{NL}}}\frac{\partial \rho_{\tiny \mbox{NL}}}{\partial x} =\frac{\gamma}{2}\tilde{x} \, ,
\end{equation}
then, the total velocity field: $v_{\tiny \mbox{NL,tot}}(x,t)$ is, like in the non-dissipative case, given by
\begin{equation}
v_{\tiny \mbox{NL,tot}}(x,t)=v_{\tiny \mbox{NL}}(x,t)+v_{D_x}(x,t)=\dot{\eta}(t)+\frac{\dot{\alpha}_{\tiny \mbox{NL}}(t)}{\alpha_{\tiny \mbox{NL}}(t)}\tilde{x} \, ,
\label{vel-tun}
\end{equation}
but, now with $\alpha_{\tiny \mbox{NL}}(t)$ replacing $\alpha_{\tiny \mbox{L}}(t)$. 

\begin{figure}[ht!]
\begin{center}
\includegraphics[width=0.6\textwidth]{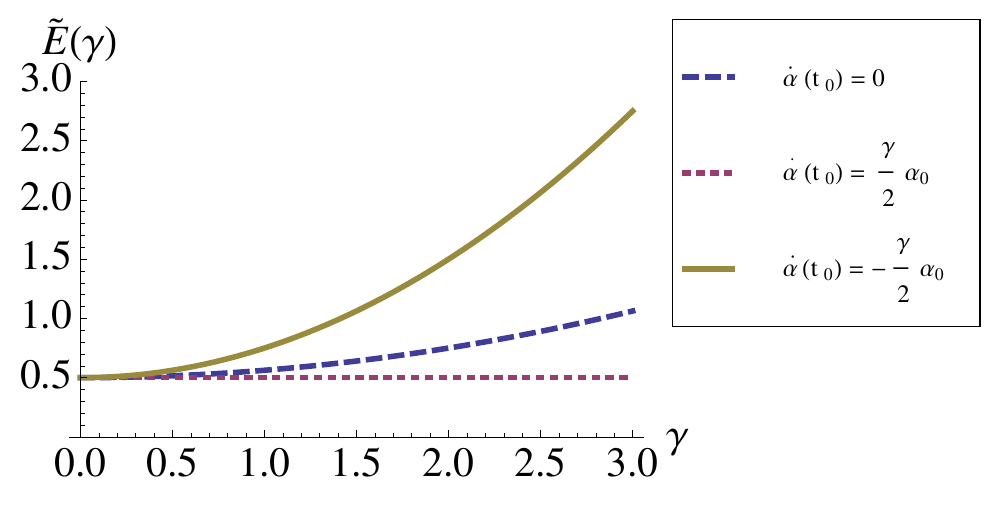}
\caption{\footnotesize The quantum contribution to the energy as a function of the environmental parameter $\gamma$, $\tilde{E}(\gamma)$, for an initial time $t_0=0$, where $\hbar=\omega_0=\alpha_0=1$ have been considered.}
\label{Fig-1.0}
\end{center}
\end{figure}

The quantum contribution to the energy for a given initial time $t_0=0$ as a function of gamma is given by the expression 
\begin{equation}
 \tilde{E}(\gamma)=\frac{ \hbar}{4}\left[ \left( \pm|\dot{\alpha}_0|-\frac{\gamma}{2}\alpha_0 \right)^2+\frac{1}{\alpha_0^2}+\omega_0^2 \, \alpha^2_0 \right]
\end{equation}
such that when $\gamma\to 0$ one gets the quantum energy for the non-dissipative case. Fig.~\ref{Fig-1.0} shows $\tilde{E}(\gamma)$ for the particular initial conditions: $(\alpha_0,|\dot{\alpha}_0|)=(\alpha_0,0)$ and $(\alpha_0,|\dot{\alpha}_0|)=(\alpha_0,\frac{\gamma}{2}\alpha_0)$. The figure displays that the quantum contribution to the energy in the non-dissipative case, $\tilde{E}(\gamma=0)$, bifurcates in three different energies as soon as the parameter $\gamma$ becomes different from zero, enhancing the importance of, and sensitivity to, the initial conditions.

In the following again, like in Section \ref{section-3}, the subscript "NL" is dropped.  Additionally we are going to describe, in the next examples, the uncertainties, correlations, tunnelling currents, and the quantum contribution to the energy (see Table~\ref{Table-1}). For this purpose we will need in each case the explicit expressions of $\xi_i(t)$ and its corresponding momenta $g_i(t)$. 
 
 
\subsection{Damped free motion}

First of all, for $\omega=0$, we obtain as solution for the damped Newtonian equation 
\begin{equation}
\langle x \rangle(t)=\eta(t)=\eta_0 +\frac{\dot{\eta}_0 }{\gamma}\left( 1- e^{-\gamma t} \right) \, ,
\end{equation}
with the initials conditions $\eta(0)\equiv \eta_0$ and $\dot{\eta}(0) \equiv\dot{\eta}_0$. These allow us to determine the time-evolution of the position and momentum mean values.

The two linear independent solutions of Eq.~(\ref{Phy-New-Eq}) with initial condition (\ref{init-cond-New}) and the corresponding momenta are 
\begin{equation}
\xi_1(t)=-\frac{2}{m\gamma}\sinh\left( \frac{\gamma}{2} t \right),\quad
\xi_2(t)=\cosh\left( \frac{\gamma}{2} t \right),
\end{equation}
\begin{equation}
g_1(t)=\cosh\left( \frac{\gamma}{2}t \right), \quad g_2(t)=-\frac{\gamma m}{2}\sinh\left( \frac{\gamma}{2}t \right).
\end{equation}
Inserting the above time-dependent functions in the expressions presented in Table~\ref{Table-1}  all the quantum mechanical properties can be determined. 

To show the sensitivity to the initial conditions, we consider the cases $(\alpha_0, |\dot{\alpha}_0|)=(\alpha_0, 0)$  and $(\alpha_0, |\dot{\alpha}_0|)=(\alpha_0,\frac{\gamma}{2}\alpha_0)$. Remember that for $|\dot{\alpha}_0| \neq 0$ we will have two posibilities for all the mentioned quantities.

\begin{figure}[htb]
\centering 
\begin{subfigure}[b]{.4\linewidth}
\centering
\includegraphics[width=0.99\textwidth]{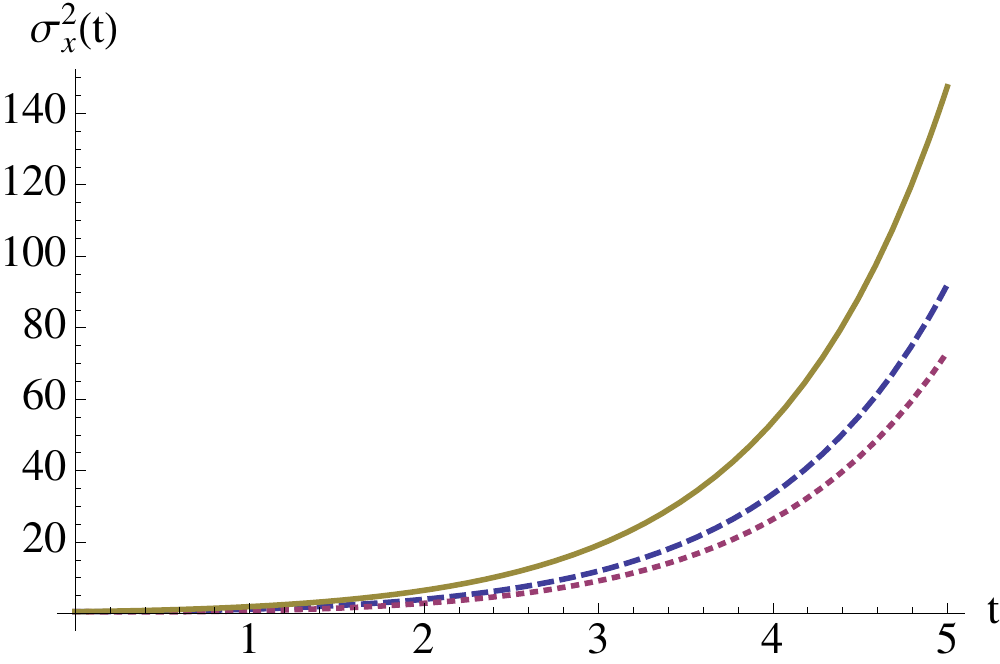} 
\caption{}
\end{subfigure}
\hskip5ex
\begin{subfigure}[b]{.4\linewidth}
\includegraphics[width=0.99\textwidth]{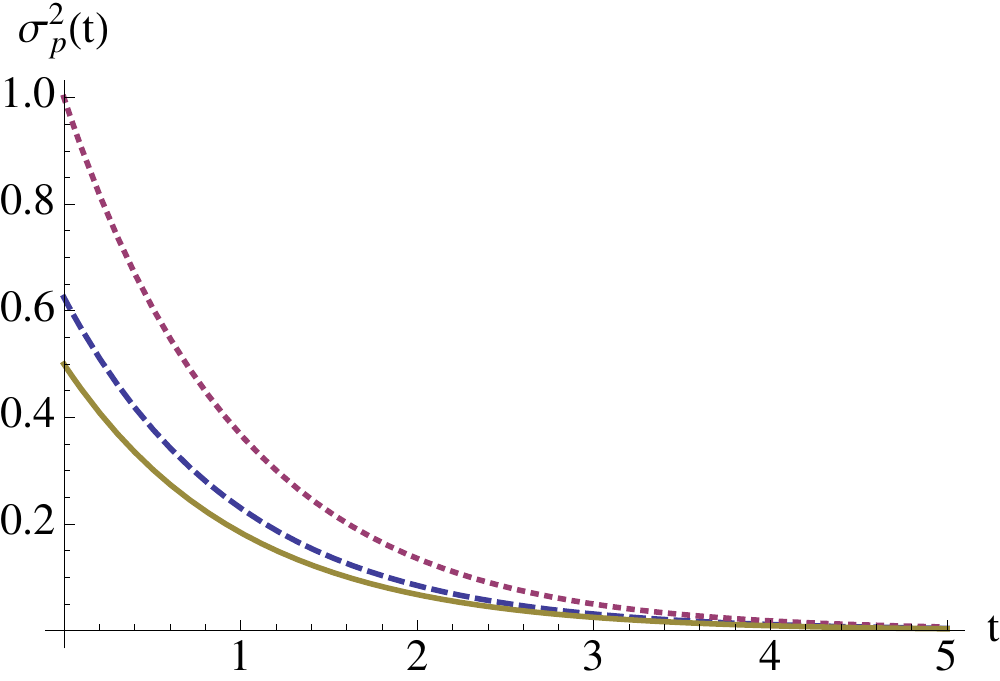} 
\caption{}
\end{subfigure}

\begin{subfigure}[b]{.4\linewidth}
\includegraphics[width=0.99\textwidth]{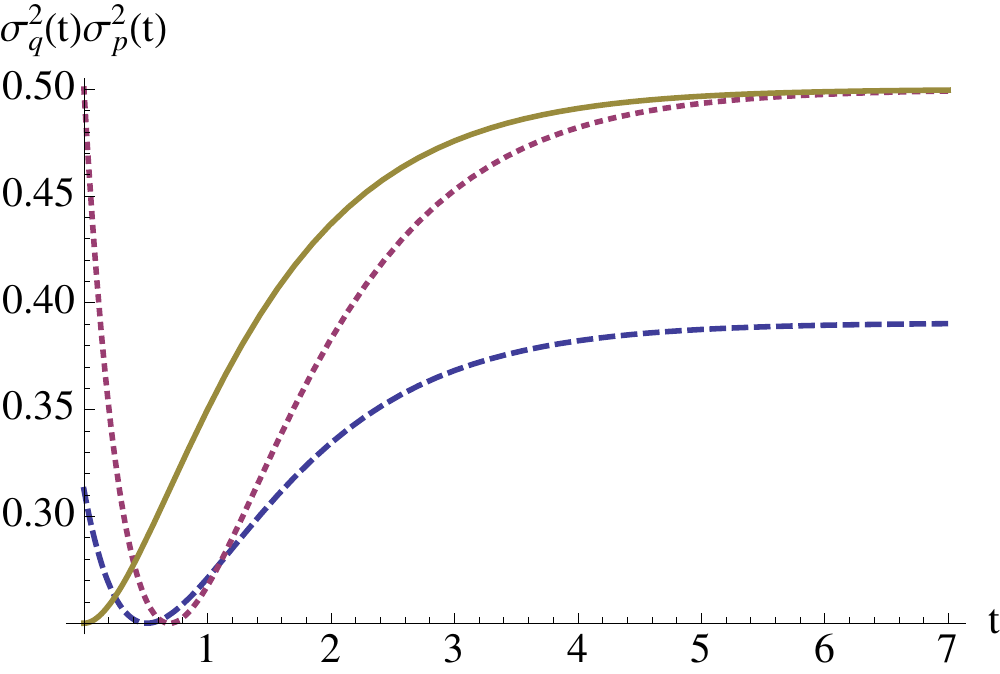}
\caption{}
\end{subfigure}
\hskip5ex
\begin{subfigure}[b]{.4\linewidth}
\includegraphics[width=1\textwidth]{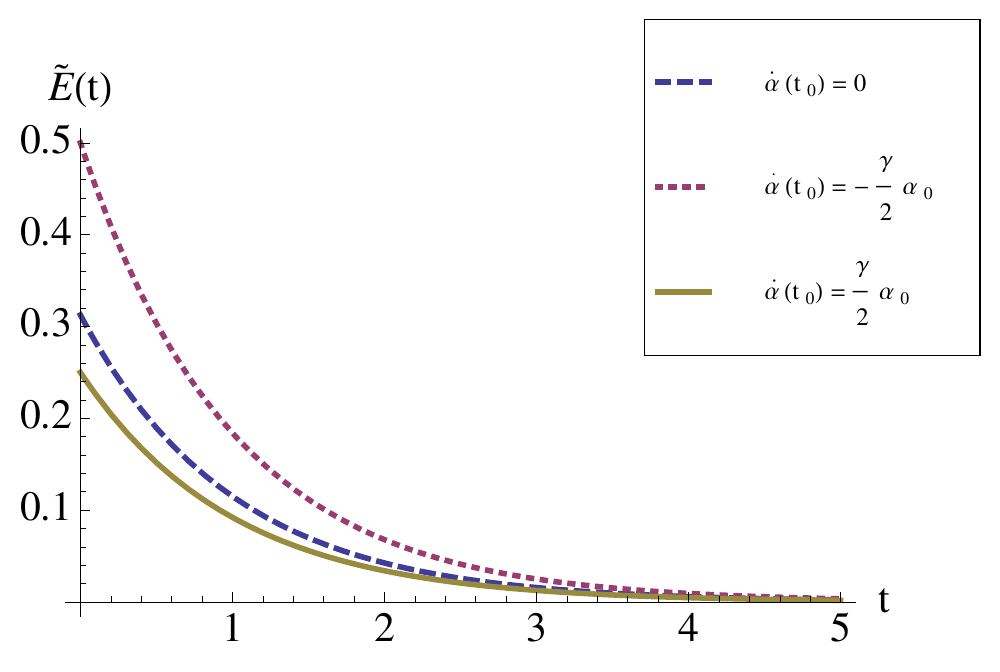}
\caption{}
\end{subfigure}
\caption{\footnotesize The time evolution of (a) position and (b) momentum uncertainties, (c) the product of the uncertainties and (d) energy. In all cases we have used $\hbar = m = \gamma =1$, with the initial condition $\alpha_0=1$.
}
\label{Fig-1}
\end{figure}

The position uncertainties, $\sigma_x^2(t)$, are increasing functions of time, see Fig.~\ref{Fig-1}(a) , whereas the momentum uncertainties, $\sigma_p^2(t)$, are decreasing time-dependent functions that go to zero as $t\to\infty$, see Fig~\ref{Fig-1}(b). However, the product of position and momentum uncertainties, $\sigma_x^2(t)\sigma_p^2(t)=\frac{\hbar^2}{4}+\sigma_{xp}^2(t)$ shown in~Fig.~\ref{Fig-1}(c), approaches a constant value depending on the choice of the initial conditions for $\alpha_0$ and $|\dot{\alpha}|_0$, but not on the ambiguity of sign mentioned before. This constant value of the product $\sigma_x^2(t)\sigma_p^2(t)$ can be determined in general through the expression 
 \begin{eqnarray}
\lim_{t\to\infty}\sigma_x^2(t)\sigma_p^2(t)&=&\frac{\hbar^2}{4}+\left(\frac{1}{m\gamma}\sigma_{p_0}^2+\sigma_{xp_0}\right)^2 \, , \nonumber\\
&=&\frac{\hbar^2}{4}\left[1+\frac{1}{\gamma^2}\left(|\dot{\alpha}_0|^2+\frac{1}{\alpha_0^2}-\frac{\gamma^2}{4}\alpha_0^2 \right)^2\right].
 \end{eqnarray}
 Thus, for the initial condition $|\dot{\alpha}_0|=\frac{\gamma}{2}\alpha_0$ this limit takes the form
  \begin{equation}
\lim_{t\to\infty}\sigma_x^2(t)\sigma_p^2(t)=\frac{\hbar^2}{4}\left[1+\left(\frac{\beta_0}{\gamma}\right)^2\right].
\label{product}
 \end{equation}
 A peculiarity of this case (that includes the minimum uncertainty WP) is that in the limit $\gamma\to\infty$, this takes the value $\frac{\hbar^2}{4}$, i.e., permanent interaction with the environment or continuous observations of the system keeps it in the (initial) state that minimizes the Heisenberg inequality, what shows similarities with the quantum Zeno effect~\cite{Degasperis, Misra}. 

For the selected initial conditions, the quantum energies decay exponentially, see Fig.~\ref{Fig-1}(d), but have different initial values. 

In general the energy gap for the two possible signs of $\dot{\alpha}(t_0)=\pm |\dot{\alpha}_0|$,  $\Delta\tilde{E}(t_0)=\tilde{E}_{-}(t_0)-\tilde{E}_{+}(t_0)$ at the initial time $t_0$ can be expressed in the form 
\begin{equation}
\Delta\tilde{E}_0=\frac{\hbar \gamma}{2}|\dot{\alpha}_0|\alpha_0=2 m D_{x_0}\frac{|\dot{\alpha}_0|}{\alpha_0},
\end{equation}
where $\alpha_0^2=\frac{2m}{\hbar}\sigma_{x_0}^2$ and $D_{x_0}=\frac{\gamma}{2}\sigma_{x_0}^2$ (following from the separation condition (\ref{Difusion-term}) of the Smoluchowski equation (\ref{Flokker-Planck-equation})) have been used. This splitting of the energy can be interpreted as a bifurcation of the energy of the non-dissipative case, $\tilde{E}=\frac{\hbar}{4}\beta_0$, into different values due to the breaking of time-reversal symmetry~\cite{Dieter-3}.

In particular for $|\dot{\alpha}_0|=\frac{\gamma}{2}\alpha_0$ the gap is given by $\Delta\tilde{E}_0=m\gamma D_{x_0}$, which depends only on the properties of the environment and the expression does not depend explicitly on $\hbar$! So, essentially the friction coefficient $\gamma$ and the diffusion coefficient $D_{x_0}$ determine the addition to the quantum mechanical energy contribution $\tilde{E}$, similar to the increase of the energy of a Brownian particle due to the stochastic force. This similarity can be taken even further assuming that at the initial time the environment was in thermal equilibrium, therefore applying the Einstein relation $D=\frac{kT}{m\gamma}$ for the diffusion coefficient $D_{x_0}$. This would lead for $\vert \dot{\alpha}_0 \vert= \frac{\gamma}{2} \, \alpha_0$ to an initial energy gap 
\begin{equation}
\Delta\tilde{E}_0=k T , 
\end{equation}
with $k$ being the Boltzmann's constant and $T$ the temperature, and in the case $|\dot{\alpha}_0|=0$ to an additional energy contribution of $\frac{1}{4}kT$. Similar considerations are also possible in the case of the damped HO, as shown below.


\subsection{Damped harmonic oscillator}

In this subsection three cases have to be discussed: under-critical damping ($\omega_0 > \frac{\gamma}{2}$), aperiodic limit ($\omega_0=\frac{\gamma}{2}$) and overdamping ($\omega_0~<~\frac{\gamma}{2}$). Thus, the classical trajectories of the mean values $\eta(t)=\langle x \rangle(t)$ for the three cases are:
\begin{eqnarray}
\omega_0 > \frac{\gamma}{2}&:& \quad \eta(t)=\left[ \eta_0 \cos{\Omega t}+\frac{1}{\Omega}\left( \frac{\gamma}{2}\eta_0+\dot{\eta}_0 \right)\sin{\Omega t} \right]e^{-\frac{\gamma t}{2}}\,  ,\\
\omega_0=\frac{\gamma}{2}&:& \quad \eta(t)=\left[\left(1+\frac{\gamma}{2} t\right)\eta_0+ \dot{\eta}_0t \right]e^{-\frac{\gamma}{2} t}\, ,\\
\omega_0 < \frac{\gamma}{2}&:& \quad \eta(t)=\left[ \eta_0 \cosh{\tilde{\Omega} t}+\frac{1}{\tilde{\Omega}}\left( \frac{\gamma}{2}\eta_0+\dot{\eta}_0 \right)\sinh{\tilde{\Omega} t} \right]e^{-\frac{\gamma t}{2}}\, ,
\end{eqnarray}
with the quantities $\Omega^2=\omega^2_0-\frac{\gamma^2}{4}$ and $\tilde{\Omega}^2=\frac{\gamma^2}{4}-\omega^2_0$. 

The two linear independent solutions $\xi_i(t)$ of Eq.~(\ref{Phy-New-Eq}) with initial condition (\ref{init-cond-New}) and their momenta $g_i(t)=-m\dot{\xi}_i(t)$ are
\begin{eqnarray}
\omega_0 > \frac{\gamma}{2}&:&\qquad\xi_1(t)=-\frac{1}{m\Omega}\sin{\Omega t}\, , \quad \xi_2(t)=\cos{\Omega t}\, ;\nonumber \\
&&\qquad g_1(t)= \cos{\Omega t} \, , \quad g_2(t)=m\Omega\sin{\Omega t}\, .\label{Ap-lim}\\
\omega_0=\frac{\gamma}{2}&:&\qquad\xi_1(t)=-\frac{t}{m}\, , \quad \xi_2(t)=1\, ; \label{Ap-lim-1} \nonumber \\
&&\qquad  g_1(t)= 1\, , \quad g_2(t)=0\, . \\
\omega_0 < \frac{\gamma}{2}&:&\qquad\xi_1(t)=-\frac{1}{m\tilde{\Omega}}\sinh{\tilde{\Omega} t}\, , \quad \xi_2(t)=\cosh{\tilde{\Omega} t}\, ; \label{Over} \nonumber \\
&&\qquad g_1(t)= \cosh{\tilde{\Omega} t}\, , \quad g_2(t)=-m\tilde{\Omega}\sinh{\tilde{\Omega} t}\, .\label{Over-1}
\end{eqnarray}
Thus, the observables presented in Table~\ref{Table-1} can be constructed inserting for each case the corresponding $\xi_i(t)$ and $g_i(t)$.

\medskip

\noindent
{\it Under-critical damping}

From the expressions in the Table~\ref{Table-1}, one can find that the uncertainties are oscillating functions of $t$ with the reduced frequency $\Omega^2=\omega_0^2-\frac{\gamma^2}{4}$. 

An interesting case that shows the influence of the initial condition $\alpha_0$ (like in the non-dissipative case) is given for the choice $\alpha_0=\frac{1}{\sqrt{\Omega}}$, which corresponds to the singular point of the Ermakov equation. In this case the uncertainties are constant and given by
\begin{eqnarray}
\sigma_x^2(t)&=&\frac{\hbar}{2m\Omega}\, ,\\
\sigma_p^2(t)&=&\frac{\hbar m \Omega}{2}\left[1 + \left(\frac{\gamma/2}{\Omega}\right)^2 \right].
\end{eqnarray}
A constant value of $\sigma_x^2(t)$ corresponds to a constant width of the WP.  We remark you that in the dissipative case $\dot{\alpha}(t)\neq 0$ does not mean $\sigma_{xp}=0$. For this case also $\sigma_{xp}(t)$ can have a constant value different from zero, i.e., 
\begin{equation}
\sigma_{x}^2(t)\sigma_p^2(t)=\frac{\hbar^2}{4}+\sigma^2_{xp}(t)=\frac{\hbar^2}{4}\left[ 1 + \left(\frac{\gamma/2}{\Omega}\right)^2 \right].
\end{equation}
For $\gamma\to 0$ this converges to the minimum uncertainty result of the coherent state of the HO with frequency $\omega_0$.

Therefore the WP has $\beta_0=\frac{\hbar}{2m\sigma^2_{x_0}}=\Omega$ which yields the quantum energy contribution 
\begin{equation}
\tilde{E}=\frac{\hbar}{4}\alpha_0^2\left[ \beta_0^2+\frac{\gamma^2}{4}\right]=\frac{\hbar \omega_0^2}{2 \Omega}
=\frac{\hbar}{2}\Omega+\frac{\hbar}{2}\frac{\gamma^2/4}{\Omega}> \frac{\hbar}{2}\omega_0,
\end{equation}
which is a constant. In particular, assuming that at the initial time the environment was in thermal equilibrium, the energy takes the form
\begin{eqnarray}
\tilde{E}=\frac{\hbar}{2}\Omega+\frac{1}{2}kT.
\end{eqnarray}
For the selection $|\dot{\alpha}_0|=\frac{\gamma}{2}\alpha_0$, i.e., WPs with TD width, the quantum contributions to the energy gap $\Delta\tilde{E}(t)$ can again be expressed in terms of the environmental parameters. 

\medskip

\noindent
{\it Aperiodic limit}

From Eq.~(\ref{Ap-lim-1}) all the quantities from Table~\ref{Table-1} can be constructed. Then, in contrast to the under-critical case,  the quantum uncertainties grow parabolic in time. However, although  the uncertainties as well as the correlation function are increasing functions of time they minimize the Robertson--Schr\"odinger uncertainty relation.

Another interesting result is that although the system is loosing its classical energy for $t \to \infty$, it regains energy from the environment via $\tilde{E}(t)$. 
This is however, not against the second law of thermodynamics that not allows a transfer of the thermal energy of the heat bath into mechanical energy of the system. In our case this energy is not transformed into classical degrees of freedom of the system, i.e., the maximum of the WP does not start oscillating with increasing amplitude, but into its quantum mechanical degrees of freedom. A better understanding of this effect needs further investigation. 

Additionally, the difference $\Delta \tilde{E}(t)$ increases linearly in time, a behaviour that is completely different from the one presented in the damped free motion, where the difference converges to zero, and the one presented in the  under-critical case, where this difference oscillates in time. 

\medskip

\noindent
{\it Overdamping}

In this case with the help of the expressions in Eq. (\ref{Over}) the quantities of Table~\ref{Table-1}  can be determined. So, due to the occurrence of the hyperbolic functions in the corresponding $\xi_i(t)$ and $g_i(t)$, these quantities are time dependent functions that diverge exponentially. Consequently, the quantum energy contributions $\tilde{E}(t)$ also increase exponentially in time, therefore, as in the aperiodic limit, the system gains energy from the environment via $\tilde{E}$ and the gap $\Delta\tilde{E}(t)$ also increases in time. 


\section{ Conclusions and perspectives}

We have extended the formalism developed in part I in order to describe dissipative quantum systems. We started by introducing the effective models of CK and expanding coordinates, which use the canonical formalism. Consequently, the SEs of these models were obtained by canonical quantization. The connection between both descriptions was established via a canonical transformation on the classical level or by a unitary transformation on the quantum level. However, in particular in the quantum mechanical case the observables must be expressed in terms of the physical variables to avoid problems such as the violation of the uncertainty principle as happened in the CK-approach.  

A different approach was considered introducing a difussion term into the continuity equation giving rise to a Smoluchowski-type equation for the probability distribution $\rho_{\tiny \mbox{NL}}(x,t)$. This yields a NLSE for the WP with well-defined physical interpretation. The connections of this approach with the effective models of CK and expanding coordinates was also establish by means of a classically non-canonical and quantum mechanically non-unitary transformation. Furthermore, the established transformations amongst effective models in Section~\ref{section-2} show that all the descriptions are equivalent in terms of the physical quantities. 

Associated to each effective model there is an Ermakov system whose solutions provides the information about the dynamics of the quantum mechanical properties.  These properties, such as the uncertainties and their correlation function, determine the quantum contribution of the energy and the tunnelling currents, as it was demonstrated in Section~\ref{section-3}. 

The existence of the Ermakov invariant and its connection with the expanding-coordinates-approach allows to use the Appendix B of part I to find the NL Ermakov solutions for the damped HO. Additionally the real and imaginary parts of the linearized Riccati equation, connected with the mean values and the uncertainties of the position and momentum operators, allow us to construct the Feynman kernel exactly in the same way as presented in part I.

The Wigner function was also constructed and, although it has the same functional form as in the non-dissipative case, it is not directly determined by the corresponding Ermakov invariant but has to be multiplied by an exponential factor.  By substitution it was shown that the Wigner function satisfies a Fokker-Planck-type equation and its marginals give us the probability distribution functions in the position and momentum representations as is expected, where the one in position space fullfills the  Smoluchowski equation.

A crucial point of our discussion, in part I and in this contribution, was the sensitivity of the evolution of the quantum system to the choices of the initial conditions.  This sensitivity is more obvious in the dissipative case due to the presence of the environmental parameter $\gamma$ as is demonstrated in Section~\ref{section-6}. In particular, we demonstrate the different behaviour of the dissipative and non dissipative systems for the same initial conditions.  Although the tunnelling currents in both cases have the same form as a function of $\alpha(t)$, the Ermakov equations for $\alpha(t)$ are different.  

We have proved that the quantum energy $\tilde{E}$ is susceptible to the initial conditions due to the presence of $\gamma$. In particular for the damped free motion and the damped HO systems it was discussed how the environmental parameter can produce a bifurcation of the quantum energy compared to the one without dissipation~(see Fig.~\ref{Fig-1.0}). An important characteristic of this effective description of dissipative systems is that, applying the Einstein relation $D=\frac{kT}{m\gamma}$ for the diffusion coefficient, i.e., assuming that at the initial time the environment is in thermodynamic equilibrium, one can express the gap of the quantum energy in terms of the temperature of the environment. 

For the damped free motion, $\omega_0=0$, all the properties of the quantum system were constructed for the initial conditions $|\dot{\alpha}_0|=0$ and $|\dot{\alpha}_0|=\frac{\gamma}{2}\alpha_0$. It has been demonstrated that the physical properties depend strongly on the election of the sign of $\dot{\alpha}_0$. Furthermore, each sign defines a different contribution to the energy, where the separation between them is depending on the environmental parameters $\gamma$ and $D_x(t)$. The product of momentum and position uncertainties in the limit $t \to \infty$ goes to a constant value. 
For the initial condition $|\dot{\alpha}_0|=\frac{\gamma}{2}\alpha_0$, this constant takes the form given in expression~(\ref{product}), which in the limit $\gamma \to \infty$ takes the value $\hbar^2/4$, a result that shows similarity with the quantum Zeno effect because the permanent interaction with the environment or continuous observation of the system keeps it in the initial state (coherent state). 

For the damped HO, we start by analyzing the behaviour of the quantum observables for $\omega_0 > \frac{\gamma}{2}$. The behaviour of the observables are similar to those reported for the HO in part I, except that  the mean values of position and momentum operators are now damped oscillating functions. 
The initial condition $ \alpha_0=1/\sqrt\Omega$ yields constant uncertainties, that in the limit $\gamma\to 0$ describes a coherent state for the HO. In contrast, for other initial conditions, the uncertainties are oscillating functions of time.  

For the aperiodic limit ($\omega_0=\frac{\gamma}{2}$) and the overdamping ($\omega_0~<~\frac{\gamma}{2}$) the quantum energy $\tilde{E}(t)$ is an increasing function of time, while the classical part of the energy goes to zero. 

The formalism developed in part I and presented in this contribution can also be extended to a TD frequency. Thus, in the third part of our contribution we will introduce how our treatment can be extended easily to cases where no analytic solutions for the classical equation of motion may exist with and without dissipation. For dissipative systems we are going to consider the environmental parameter $\gamma(t)$ as a time-dependent function. Particularly, we are interested in the evolution where the environmental parameter or the TD frequency change suddenly, linearly, and adiabatically. 

\label{section-7}


\subsection*{Acknowledgments}
This work was partially supported by CONACyT-M\'exico (under project 238494 and 152574).


\appendix

\section{Dynamics of the quantum uncertainties}

\label{Appendix-A}

For the dissipative system expressed in the physical variables, the uncertainties of position and momentum and their correlation satisfy the set of differential equations given in Eqs.~(\ref{dispersion-1}) to (\ref{dispersion-3}).

This system has the invariant $I_{SR}=\sigma_x^2(t)\sigma_p^2(t)-\sigma_{xp}^2(t)=\frac{\hbar^2}{4}$. In order to solve the system of differential equations, it can be equivalently expressed as a third-order ordinary differential equation for the position uncertainty, 
\begin{equation}
\frac{d^3\sigma^2_x(t)}{dt^3}+4\Omega^2(t)\frac{d\sigma_x^2(t)}{dt}+4\Omega(t)\frac{d\Omega(t)}{dt}\sigma^2_x(t)=0\, ,
\label{TODE}
\end{equation}
where $\Omega^2(t)=\omega^2(t)-\frac{\gamma^2}{4}$. This equation is the same as the one obtained in the non-dissipative case, Eq. (A.4) of part I, just replacing $\omega(t)$ by $\Omega(t)$; therefore, we propose the same solution, i.e., 
\begin{equation}
\sigma_x^2(t)=\frac{\hbar}{2m}\alpha^2(t)\, .
\label{sx-sol}
\end{equation}
Thus, inserting this solution into (\ref{TODE}), it shows that it is solution only if $\alpha(t)$ satisfies the Ermakov equation (\ref{Damped-Ermakov-equation}). Substituting solution (\ref{sx-sol}) into (\ref{dispersion-1}) and solving for $\sigma_{xp}(t)$ one obtains that
\begin{equation}
\sigma_{xp}(t)=\frac{\hbar}{2}\alpha(t)\left[ \dot{\alpha}(t)-\frac{\gamma}{2}\alpha(t) \right]\, .
\label{sxp-sol}
\end{equation}
Inserting now (\ref{sx-sol}), (\ref{sxp-sol}) into (\ref{dispersion-3}), and solving for $\sigma_p^2(t)$ one finds that 
\begin{equation}
\sigma^2_p(t)=\frac{m \hbar}{2} \left[ \left( \dot{\alpha}(t)-\frac{\gamma}{2}\alpha(t) \right)^2+\frac{1}{\alpha^2(t)} \right]\, . \label{sp-sol}
\end{equation}

Because the position uncertainty is proportional to $\alpha^2(t)$, it can directly be obtained from Eq.~(\ref{Erm-New}) for given $\alpha_0$ and $|\dot{\alpha}_0|$, once the $\xi_1(t)$ and $\xi_2(t)$ solutions of Eq.~(\ref{Phy-New-Eq}) with initial conditions (\ref{init-cond-New}) are determined. Thus, the position uncertainty in terms of the $\xi(t)$ is given by
\begin{equation}
\sigma_x^2(t)=\frac{\hbar}{2 m}\left[ m^2\beta_0\xi_1^2(t)+\left(m|\dot{\alpha}_0|\xi_1(t)\mp\alpha_0\xi_2(t) \right)^2 \right]
\end{equation}
with the frequency-type quantity $\beta_0=\frac{1}{\alpha_0^2}$. 

On the other hand, the uncertainty $\sigma_p^2(t)$ and the correlation function $\sigma_{xp}(t)$ can be constructed from Eq.~(\ref{Erm-New}) and the relations
\begin{eqnarray}
\dot{\alpha}_{\mp}^2(t)+\frac{1}{\alpha_{\mp}^2(t)}&=&\frac{\hbar}{m}\left[ A g_1^2(t)+B g_2^2(t)\mp 2C g_1(t)g_2(t)\right],\label{cons-sp}\\
\dot{\alpha}_{\mp}(t)\alpha_{\mp}(t)&=&-\hbar\left[A\xi_1(t)g_1(t)+B\xi_2(t)g_2(t)\mp C( \xi_1(t)g_2(t)+\xi_2(t)g_1(t))\right]\, ,\qquad
\label{cons-sxp}
\end{eqnarray} 
where the $g_i(t)$ are, up to a constant, the time-derivative of $\xi_i(t)$, $g_i(t)=-m\dot{\xi}_i(t)$. Therefore, general expressions for $\sigma_p^2(t)$ and $\sigma_{xp}(t)$ in term of $\xi_i(t)$, $g_i(t)$ and the initial conditions $\alpha_0$ and $|\dot{\alpha}_0|$ are attained inserting (\ref{Erm-New}), (\ref{cons-sp}) and (\ref{cons-sxp}) into (\ref{sp-sol}) and (\ref{sxp-sol}), respectively, thus
\begin{eqnarray}
\sigma_p^2(t)&=&\frac{\hbar m}{2}\left[\chi_1^2(t)\beta_0+\left(\chi_1(t)|\dot{\alpha}_0|\mp\chi_2(t)\frac{\alpha_0}{m} \right)^2\right]\, ,\\
\sigma_{xp}(t)&=&\frac{\hbar}{2}\Bigg[ \left( m\left[|\dot{\alpha}_0|^2+\frac{1}{\alpha_0^2}\right]\mp|\dot{\alpha}_0|\alpha_0 \right)\chi_1(t)\xi_1(t)\nonumber\\
        &&+\left(\frac{\alpha_0^2}{m}\mp|\dot{\alpha}_0|\alpha_0 \right)\chi_2(t)\xi_2(t)\Bigg]\, ,
\end{eqnarray}
with $\chi_i(t)=-g_i(t)-m\frac{\gamma}{2} \xi_i(t)$, $i=1,2$. 


\end{document}